\documentclass[sigconf]{acmart}

\usepackage[utf8]{inputenc}
\usepackage[T1]{fontenc}
\usepackage{url}

\usepackage{graphicx}

\usepackage{booktabs} % For formal tables
\usepackage{xspace}
\usepackage{dsfont}
\usepackage{siunitx}
\usepackage{subcaption}
\usepackage{version}
\usepackage{amsmath}
\usepackage{amssymb}

\usepackage[symbol]{footmisc}

\excludeversion{commentaires}
\excludeversion{commentaires-imp}
\excludeversion{longversion}
\includeversion{journal}
\excludeversion{conference}
\excludeversion{shortconference}

\usepackage{algorithm}
\usepackage[noend]{algpseudocode}
\usepackage{color}
\usepackage{xcolor}

\newcommand\numberthis{\addtocounter{equation}{1}\tag{\theequation}}

\newcommand{\rejectforwg}[1]{}

\newcommand{\acceptforwg}[1]{#1}

\newcommand{\fred}[1]{{\color{blue} [Fred: #1]}}
\newcommand{\fredvu}[1]{}
\newcommand{\thib}[1]{{\color{purple} #1}}
\newcommand{\thibvu}[1]{}

\newcommand{\giovannivu}[1]{}
\def\nitbf{\noindent\textbf}
\def\niem{\noindent\emph}
\def\ms{\medskip}

\def\cc{clustering coefficient\xspace}

\def\icc{interest clustering coefficient\xspace}
\def\Icc{Interest clustering coefficient\xspace}
\def\iccs{{\tt icc}\xspace}
\def\ts{Twitter Snapshot\xspace}
\def\tss{TS\xspace}

\def\ccUT{undirected clustering coefficient\xspace}
\def\ccUTs{ucc\xspace}
\def\ccMUT{mutual clustering coefficient\xspace}
\def\ccMUTs{mcc\xspace}
\def\ccTT{transitive clustering coefficient\xspace}
\def\ccTTs{tcc\xspace}
\def\ccCT{cyclic clustering coefficient\xspace}
\def\ccCTs{ccc\xspace}

\def\ra{\rightarrow}

\def\dmmax{d^{-}_{\max}}
\def\dpmax{d^{+}_{\max}}

\def\Pplus{P^{+}\xspace}
\def\Pmoins{P^{-}\xspace}
\def\Cplus{C^{+}\xspace}
\def\Cmoins{C^{-}\xspace}
\def\aplus{\alpha^{+}\xspace}
\def\amoins{\alpha^{-}\xspace}

\def\adjacencyMemory{\SI{92}{GB}\xspace}
\def\machineMemory{\SI{192}{GB}\xspace}

\graphicspath{{figures/}}

\newcommand{\delete}[1]{}

\setcopyright{none}
\renewcommand\footnotetextcopyrightpermission[1]{} % removes footnote with conference information in first column
\makeatletter
\renewcommand\@formatdoi[1]{\ignorespaces}
\makeatother

\acmConference[]{}{}{}
\title{Interest Clustering Coefficient: a New Metric for Directed Networks like Twitter}

\author{Thibaud Trolliet}
\affiliation{%
\institution{Inria Sophia-Antipolis}
\state{France}
}
\email{thibaud.trolliet@inria.fr}

\author{Nathann Cohen}
\affiliation{
\institution{Université Côte d'Azur/CNRS}
\state{France}
}
\email{nathann.cohen@cnrs.fr}

\author{Frédéric Giroire}
\affiliation{
\institution{Université Côte d'Azur/CNRS}
\state{France}
}
\email{frederic.giroire@cnrs.fr}

\author{Luc Hogie}
\affiliation{
\institution{Université Côte d'Azur/CNRS}
\state{France}
}
\email{luc.hogie@cnrs.fr}

\author{Stéphane Pérennes}
\affiliation{
\institution{Université Côte d'Azur/CNRS}
\state{France}
}
\email{stephane.perennes@cnrs.fr}


\begin{abstract}
We study here the clustering of {\em directed} social graphs. The clustering coefficient has been introduced to capture the social phenomena that a friend of a friend tends to be my friend. This metric has been widely studied and has shown to be of great interest to describe the characteristics of a social graph. In fact, the clustering coefficient is adapted for a graph in which the links are undirected, such as friendship links (Facebook) or professional links (LinkedIn). For a graph in which links are directed from a source of information to a consumer of information, it is no more adequate. We show that former studies have missed much of the information contained in the directed part of such graphs. 
We thus introduce a new metric to measure the clustering of a directed social graph with interest links, namely the \icc. We compute it (exactly and using sampling methods) on a very large social graph, a Twitter snapshot with 505 million users and 23 billion links.
We additionally provide the values of the formerly introduced directed and undirected metrics, a first on such a large snapshot. We exhibit that the \icc is larger than classic directed clustering coefficients introduced in the literature. 
This shows the relevancy of the metric to capture the informational aspects of directed graphs. 
\end{abstract}

\begin{document}

%\mainmatter              % start of a contribution

\maketitle

%	\author{Ben Trovato}
%	\authornote{Dr.~Trovato insisted his name be first.}
%	\orcid{1234-5678-9012}
%	\affiliation{%
%		\institution{Institute for Clarity in Documentation}
%		\streetaddress{P.O. Box 1212}
%		\city{Dublin}
%		\state{Ohio}
%		\postcode{43017-6221}
%	}
%	\email{trovato@corporation.com}

%	\author{John Smith}
%	\affiliation{\institution{The Th{\o}rv{\"a}ld Group}}
%	\email{jsmith@affiliation.org}
		
%	% The default list of authors is too long for headers.
%	\renewcommand{\shortauthors}{B. Trovato et al.}
	
	%	
%	\keywords{ACM proceedings, \LaTeX, text tagging}
	
%	\author{Thibaud Trolliet, Nathann Cohen, Fr\'ed\'eric Giroire, Luc Hogie, St\'ephane P\'erennes}

\section{Introduction}

%\vspace{-0.1cm}
Networks appear in a large number of complex systems, whether they are social, biological, economical or technological. Examples include neuronal networks, the Internet, financial transactions, online social networks, ...
Most ``real-world'' networks exhibit some properties that are not due to chance and that are really different from random networks or regular lattices. \giovannivu{Too vague. Want to stress that are neither fully ordered, neither fully random?} In this paper, we focus on the study of the clustering coefficient of social networks. Nodes in a network tend to form highly connected neighborhoods. This tendency can be measured by the clustering coefficient. It is classically defined for undirected networks as three times the number of triangles divided by the number of open triangles (formed by two incident edges).\rejectforwg{\footnote{Note that both a local and a global definition exist: we provided here the definition of the global clustering coefficient, see Section~\ref{sec:related-work}.}}
%\[
%CC = 3 \times \frac{\text{\# closed triangles in the graph}}{\text{\# connected triplets of vertices in the graph}}. 
%\]
This clustering coefficient had been computed in many social networks and had been observed as much higher than what randomness would give. 
Triangles thus are of crucial interest to understand ``real world'' networks. 

However, a large quantity of those networks are in fact directed (e.g. the web, online social networks like Instagram, financial transactions).
% \fred{However, a large number of these networks are in fact directed. If a friendship relationship goes (hopefully) both ways, this is not the case for a large number of relation types, e.g., interests such as an Internet link, a financial transaction, a following in Twitter, etc. In all these cases, the link between two persons or nodes in the network is not as simple as a single universal link. It may be directed from Alice to Bob, or from Bob to Alice, or exist in both directions.}
%It is for instance the case of Twitter, the fifth most used social network in 2019 \thib{in term of usage penetration rate}~\cite{Statista} \thib{and with 139 million daily active users}, making it one of the largest and most influential social networks. 
It is for instance the case of Twitter, one of the largest and most influential social networks with 126 million daily active users~\cite{washington-post}. In Twitter, a person can follow someone she is interested in; the resulting graph, where there is a link $u \rightarrow v$ if the account associated to the node $u$ followed the account associated to the node $v$, is thus directed. In this study, we used as main dataset the snapshot of Twitter (\tss in short) extracted by Gabielkov et al. as explained in~\cite{gabielkov2014studying} and made available by the authors. The \tss has around 505 million nodes and 23 billion arcs, making it one of the biggest snapshots of a social network available today.

The classic definition of the clustering coefficient cannot be directly applied on directed graphs. This is why most of the studies computed it on the so-called {\em mutual graph}, as defined by Myers \& al. in~\cite{myers2014information}, i.e., on the subgraph built with only the bidirectional links. 
\acceptforwg{We call {\em \ccMUT (\ccMUTs for short)}  the clustering coefficient associated with this graph.}
We computed this coefficient in the \tss, using both exact and approximated methods. 
We find a value for the \ccMUTs of 10,7\%. This is a high value, of the same order than the ones found in other web social networks. 
 % TBA? Note that we expected a similar result since we believe bidirectional edges represents for the main part the social links - if two people are friends, they are likely to follow each other. 

However, this classical way to operate {\em leaves out 2/3 of the graph!} Indeed, the bidirectional edges only represents 35\% of the edges of the \tss. 
% TBA? Indeed, Twitter is known to be a blend between a social and information media~\cite{myers2014information,kwak2010twitter}; hence most of those links are unidirectional, corresponding to its information aspect. 
%such a classical computation ignores/leaves out/leaves aside 
A way to avoid it is to consider all links as undirected \thibvu{preciser que si double lien on n'en garde qu'un ou pas ?} 
and to compute the clustering coefficient of the obtained undirected graph. 
\acceptforwg{We call {\em \ccUT (\ccUTs for short)} the corresponding computed coefficient. }
Such a computation in the \tss gives a value of \ccUTs of only 0.11\%. This is way lower than what was found in most undirected social networks.
%graphs with interesting clustering coefficient \textcolor{red}{(+ préciser que c'est aussi que ca n'a pas vraiment de sens de faire ce cc ?)}. 
It is thus a necessity to introduce specific clustering coefficients for the directed graphs. 
More generally, when analyzing any directed datasets, it is of crucial importance to take into account the information contained in its directed part in the most adequate way.

\begin{figure*}
\centering
\begin{subfigure}[b]{0.45\textwidth}
%\qquad
%\includegraphics[width=.40\columnwidth]{CC_und}
%\hfill
%\includegraphics[width=.40\columnwidth]{CC_und_pot}
\centering
\includegraphics[width=0.22\textwidth]{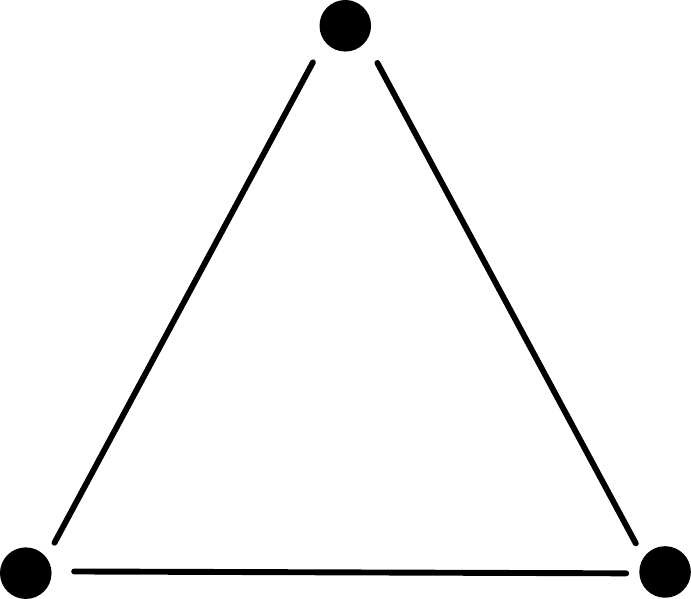}
\hspace{1.8cm}
\includegraphics[width=0.22\textwidth]{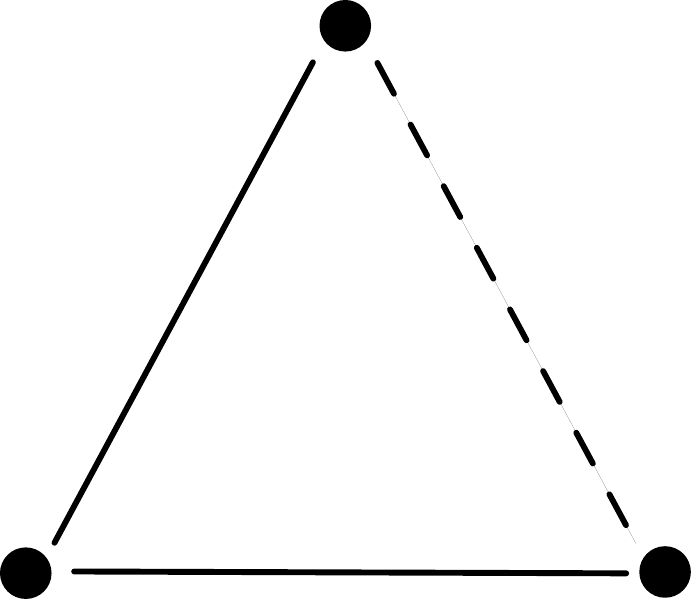}
%\qquad
\caption{Undirected triangles.
\label{fig:triangles-undirected}}
\end{subfigure}
\hfill
\begin{subfigure}[b]{0.45\textwidth}
%\qquad
%\includegraphics[width=.40\columnwidth]{CCT}
%\hfill
%\includegraphics[width=.40\columnwidth]{CCT_pot}
\centering
\includegraphics[width=0.22\textwidth]{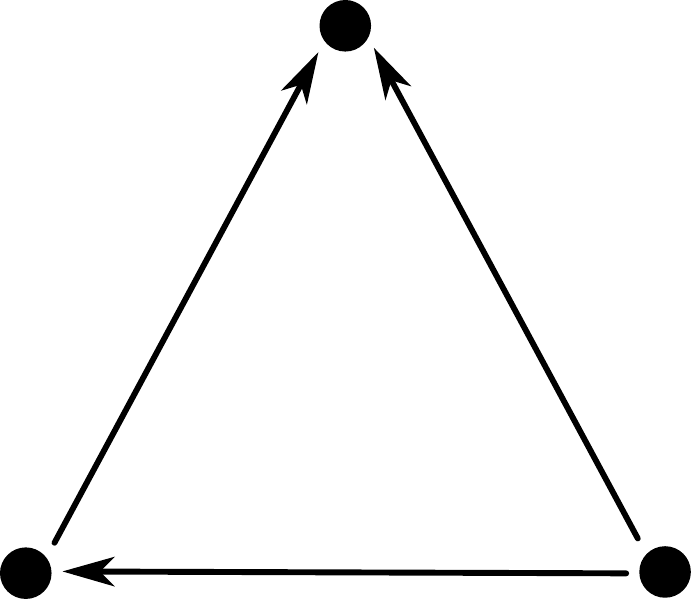}
\hspace{1.8cm}
\includegraphics[width=0.22\textwidth]{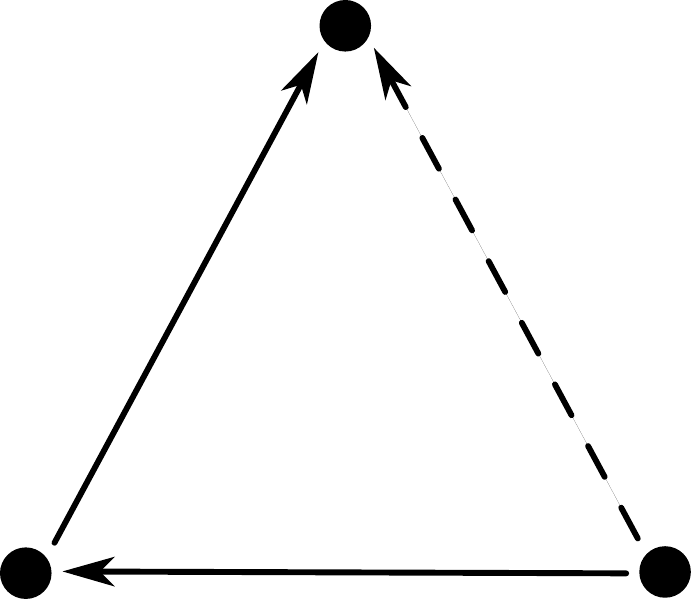}
%\qquad
\caption{Transitive directed triangles.
\label{fig:triangles-transitive}}
\end{subfigure}
\hfill
\begin{subfigure}[b]{0.45\textwidth}
%\qquad
%\includegraphics[width=.40\columnwidth]{CCC}
%\hfill
%\includegraphics[width=.40\columnwidth]{CCC_pot}
\centering
\includegraphics[width=.22\textwidth]{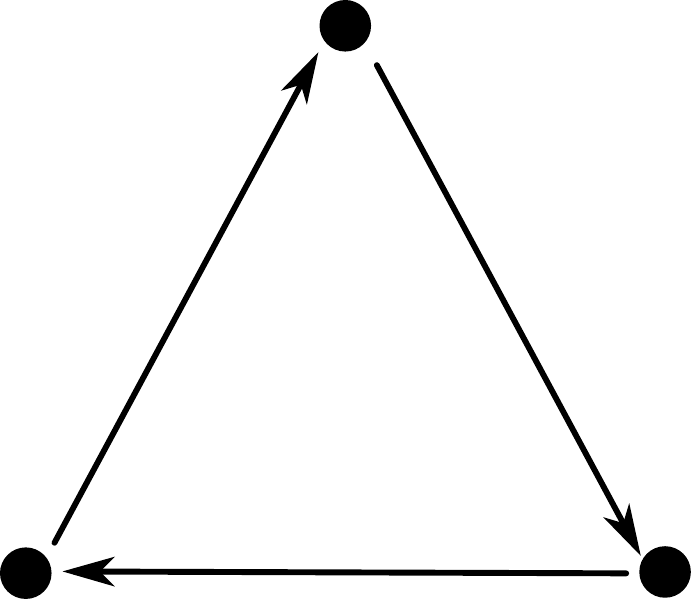}
\hspace{1.8cm}
\includegraphics[width=.22\textwidth]{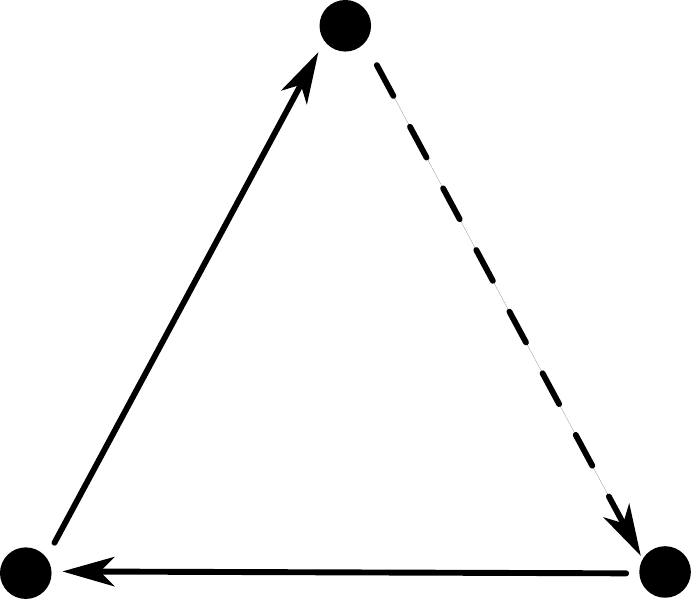}
%\qquad
\caption{Cyclic directed triangles.
\label{fig:triangles-cyclic}}
\end{subfigure}
%\vspace{-.4cm}
\hfill
\begin{subfigure}[b]{0.45\textwidth}
%\qquad
%\includegraphics[width=.35\columnwidth]{K22}
\centering
\includegraphics[width=.22\textwidth]{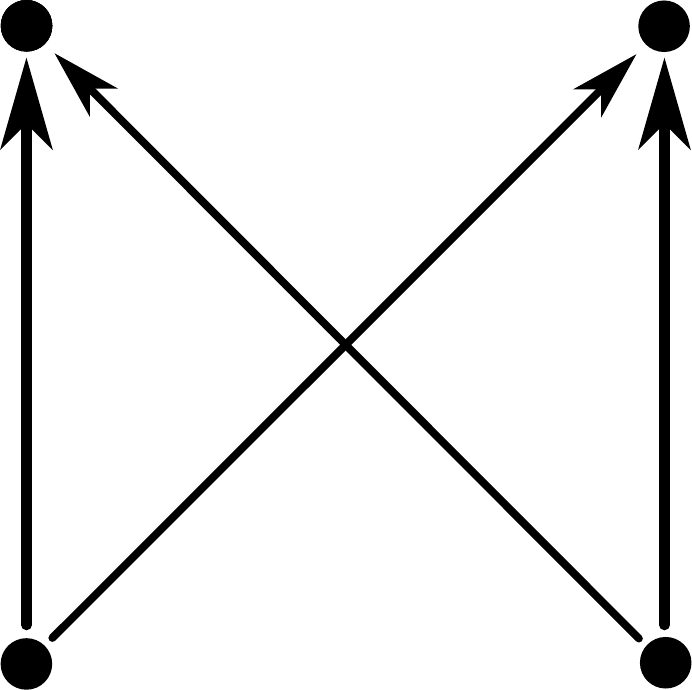}
%\hfill
\hspace{1.8cm}
\includegraphics[width=.22\textwidth]{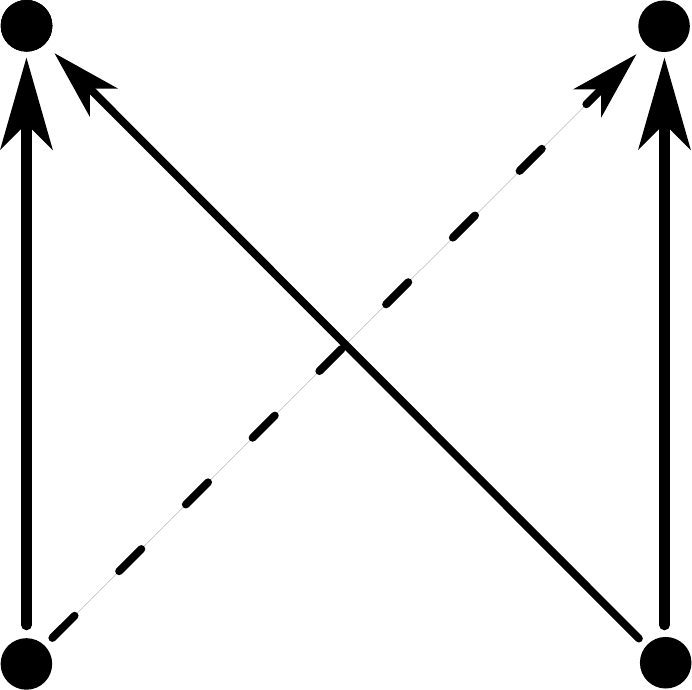}
%\qquad
\caption{K22s.}
\label{fig:K22}
\end{subfigure}
\caption{Closed (left) and open (right) undirected and directed triangles and K22s.
\label{fig:triangles}}
%\vspace{-0.6cm}
\end{figure*}

A first way to do that is to look at the different ways to form triangles with directed edges. Fagiolo computed the expected values of clustering coefficients considering directed triangles for random graphs in~\cite{fagiolo2007clustering} and illustrated his method on empirical data on world-trade flows. 
%%Fagiolo computed the expected values of clustering coefficients considering directed triangles for random graphs in~\cite{fagiolo2007clustering}. 
There are two possible orientations of triangles: transitive and cyclic triangles, 
see Figures~\ref{fig:triangles-transitive} and~\ref{fig:triangles-cyclic}. 
%see Figure~\ref{fig:triangles}. 
Each type of triangles corresponds to a directed clustering coefficient%, respectively the {\em \ccTT{} (\ccTTs in short)} and the {\em \ccCT{} (\ccCTs in short)}. 
\acceptforwg{
:
\begin{itemize}
	\item the \textbf{\textit{\ccTT{}}} (\textbf{\textit{\ccTTs}} in short), defined as:
	\[
	\ccTTs = \frac{\text{\# transitive triangles}}{\text{\# open transitive triangles}}, 
	\]
	\item the \textbf{\textit{\ccCT{}}} (\textbf{\textit{\ccCTs}} in short), defined as:
	\[
	\ccCTs = \frac{3\cdot \text{\# cyclic triangles}}{\text{\# open transitive triangles}}. 
	\]
\end{itemize}
}
We computed both coefficients for the snapshot, obtaining $\ccTTs=1.9\%$ and $\ccCTs=1.7\%$. 
%\textcolor{red}{Pour les cycliques, parler du sans bidirectional.}. The high value of the $\ccTTs$ is a direct consequence of the Twitter's retweet system \thib{(and more generally of the word of mouth system)}: if Alice follows Bob and Bob follows Carol, Bob will retweet Carol's tweets and Alice will see those retweets, increasing the probability for Alice to follow Carol and thus to close the transitive triangle. 
However, note that a large part of the transitive and cyclic triangles comes from bidirectional triangles. When removing them, we arrive to values of $\ccTTs=0.51\%$ and $\ccCTs=0.24\%$. 

We believe those metrics miss an essential aspect of the Twitter graph: 
%Twitter is known to be a blend between a social and information media~\cite{myers2014information,kwak2010twitter}. 
while the clustering coefficient were defined to represent the social cliques between people, it is not adequate to capture the information aspect of Twitter, known to be both a social and information media~\cite{kwak2010twitter,myers2014information}.
In this work, we go one step further in the way directed relationships are modeled. We argue that in directed networks, {\em the best way to define a relation or similarity between two individuals (Bob and Alice) is not always by a direct link, but by a common interest}, that is, two links towards the same node (e.g., Bob $\ra$ Carol and Alice $\ra$ Carol). Indeed, when discussing interests, consider two nodes having similar interests. Apart from being friends, these two nodes do not have any reason to be directly connected. However, they would tend to be connected to the same out-neighbors. We exploit this to study a new notion of connections in directed networks and the new naturally associated clustering coefficient, which we name {\em \icc,  or \iccs} in short, and define as follows:
%clustering coefficient of interest or K22 clustering coefficient. It has the following definition: 
%We define the \icc,  or \iccs in short, as follows
\[
\iccs = \frac{4\cdot \text{\# K22s}}{\text{\# open K22s}}, 
%\iccs = \frac{4\cdot \text{number of K22s}}{\text{number of open (connected) K22s}}, 
\]
where a K22\rejectforwg{\footnote[1]{The name comes from Graph Theory. A $K_{m,n}$ is a complete bipartite graph $G=(V_1\cup V_2,E)$ with partitions of sizes $|V_1|=m$ and $|V_2|=n$. We consider in this paper a directed version of a $K_{2,2}$.}} is defined as a set of four nodes in which two of them follow the two others, and an open K22 is a K22 with a missing link, see Figure~\ref{fig:K22}. 
We computed the \iccs on the Twitter snapshot, obtaining $\iccs=3.6\%$ ($3.1\%$ when removing the bidirectional structures). 
%As a measure of comparison, we computed the directed triangle metrics. When considering a directed graph, two orientations of triangles exist, forming transitive and cyclic triangles, see Figure~\ref{fig:triangles}. Each type of triangles corresponds to a directed clustering coefficient, respectively the \ccTT{} (\ccTTs in short) and the \ccCT{} (\ccCTs in short). We computed both coefficients for the snapshot, obtaining $\ccTTs=1.9\%$ and $\ccCTs=1.7\%$. Both are lower than the \icc. 
%If keeping the directed edges, several triangles can be defined. The most interesting one is . It defines what is called the transitivity clustering coefficient in~\cite{}. We computed this coefficient for Twitter... We see
%This shows the strong interest of our metric when dealing with interest links. 
This value, an order of magnitude higher than the previous clustering coefficients computed on the non bidirectional directed graph, confirm the interest of this metric. If the clustering coefficient of triangles are good metrics to capture the social aspect of a graph, the \icc is a good metric to capture the informational aspect. 

\rejectforwg{Using other datasets than the \ts, we will see that using all those clustering coefficients together enables a better understanding of the studied graphs.}

%We exhibit that the numbers of closed and open K22s respectively are $\num{2.56e16}$ and $\num{3.14e18}$, leading to an  \icc of Twitter of  $0.0326$. This means that, if two people have a common interest, there is a 3.26\% probability that, if one of them has another interest, the second one also shares the interest.

\;
\noindent In summary, our contributions are the following: \giovannivu{This list is heterogeneous: some bullets summarize what said before, others say more than anticipate above, one others some new contributions appear.}
%\vspace{-0.4cm}
\begin{itemize}
	\item We define a new clustering coefficient for graphs with interest links. 
	\item We succeeded in computing it, both exactly and using sampling methods, for a snapshot of Twitter with 505 million nodes and 23 billion edges. \rejectforwg{Note that we encountered several difficulties: the out-adjacency table alone used \adjacencyMemory of RAM. Moreover, for approximated methods, if triangles can be easily uniformly sampled, it is not the case for K22s and open K22s.} 
	\item We additionally provide the values of the directed and undirected clustering coefficients previously defined in the literature. We believe this is the first time that such coefficients are computed exactly for a large {\em directed} online social network. 
	\item We compute this new metric as much as the previous ones on other directed datasets to highlight the differences and interests of the different metrics. \rejectforwg{We show that their compared values give important indications about the use and structure of the datasets.}
    \item We then propose a new random graph model to obtain random directed graphs with a high \icc. We prove this model follows power-law in- and out-degree distributions, and analyse the \icc value by simulation.
    \item Lastly, we discuss the usage of this new metric for link recommendation. The principle is to recommend links closing a large number of K22s (instead, classically, of triangles). We discuss the strengths/weaknesses of this method for a set of Twitter users. 
\end{itemize}
The paper is organized as follows. We first discuss related work in Section~\ref{sec:related-work}. 
\rejectforwg{We then define the clustering coefficients for directed graphs and present the Twitter snapshot we used for the experiments in Section~\ref{sec:preliminaries}.} 
In Section~\ref{sec:computing}, we present the algorithms we used to compute the values of the \icc, both exactly and by sampling. 
We discuss the results on the clustering coefficients of Twitter in Section~\ref{sec:results}, and of other directed datasets in Section~\ref{sec:Other-Datasets}. 
In section~\ref{sec:New_Model}, we propose and study a preferential attachment model providing a high \icc. Lastly, we discuss the use of \icc for link recommendation in Section~\ref{sec:recommendations}.

% Note that the mutual graph capture the social\footnote{Note that we use the polysemous term ``social'' with two different meanings in the paper: First, in a social link (or friendship or work relations) opposed to an interest link (or information). Second, social as part of a society like in a social network, which encompasses both social and interest links.} component of the network more than the undirected graph (which keeps both interest and social links). Indeed, most friendship links are bidirectional. We should thus find again the ``friends of my friends are my friends'' phenomena in the mutual graph. Hence, we expect a high value of the \ccMUT.  

%\thib{FIGURE HERE}

%\vspace{-.4cm}

\section{Related Work}
\label{sec:related-work}

%\vspace{-0.05cm}

\nitbf{Complex networks.} Even if the study of complex networks is an old field~\cite{strogatz2001exploring}, it keeps receiving a lot of attention from the research community. 
%\thib{it has received a lot of attention from the research community in the last decades (et enlever ce qu'il y a apres)} a lot of research has been done on it in the last years \giovanni{"It keeps receiving a lot of attention from the research community".}. 
The reason for this is twofold. First, a great number of very large practical systems emerged recently can be seen as complex networks, in particular online social media networks, see~\cite{lu2011link} for a survey. Second,  with the development of big data analysis, entrepreneurs, analysts or researchers have new tools to study those huge amounts of data. 
%\giovannivu{Ce qu'il y a avant : weak, too vague. }
%\fredvu{maybe add references}
%\fredvu{in general, look at references from the web conf/WWW}
%The study of complex networks has shown that, whatever the field we are looking at, those networks exhibit some properties that are really different from random networks or regular latices.
Complex networks often exhibit common properties, like small diameter~\cite{albert2002statistical}, small average distance~\cite{watts1998collective,backstrom2012four,leskovec2008planetary}, heavy tail degree distributions~\cite{clauset2009power,leskovec2008planetary}, high clustering~\cite{watts1998collective}, communities~\cite{ugander2011anatomy}, etc.

% \thib{Pour répondre aux deux parties de Giovanni qui suivent, remplacer toute la suite par : "Complex networks often exhibit common properties, like small diameter~\cite{albert2002statistical}, small average distance~\cite{watts1998collective,backstrom2012four,leskovec2008planetary}, heavy tail degree distributions~\cite{clauset2009power,leskovec2008planetary}, high clustering~\cite{watts1998collective}, communities~\cite{ugander2011anatomy}, etc." (et ne pas parler du fait que different que l'aléatoire, mais pas grave car on en parle ensuite pour le cc dans la partie suivante).}
% Complex networks exhibit some properties that are different from random networks or regular lattices. \giovanni{A definition is missing.}
% In particular, they often exhibit small diameter~\cite{albert2002statistical}, small average distance~\cite{watts1998collective,backstrom2012four,leskovec2008planetary}, heavy tail degree distributions~\cite{clauset2009power,leskovec2008planetary},  \giovanni{All these before can be captured by random graphs. For high clustering see Wats and Strogatz model.} 
% high clustering~\cite{watts1998collective}, communities~\cite{ugander2011anatomy}, etc. 

%, the study of performance of some neural networks~\cite{kim2004performance}, finding communities~\cite{} \thib{(peut-etre "Defining and identifying communities in networks" ?)}, finding hidden thematique structure~\cite{},

%\begin{conference}
%\ms
%\end{conference}
\;

\nitbf{Clustering coefficient.} Among those properties, the clustering coefficient shows that, when two people know each other, there is a high probability that those people have common friends. The clustering coefficient has numerous important applications, such as spam detection~\cite{boykin2005leveraging}, link recommendation~\cite{silva2010graph,chen2009make}, information spread~\cite{granovetter1977strength}, study of biased network samples~\cite{newman2003ego}, performance of some neural networks~\cite{kim2004performance}  \giovannivu{what is the application ?}, etc. 
%ego-centered sampled networks~\cite{newman2003ego}, etc. 
There are different definitions of the clustering coefficient. 
\begin{conference}
The {\em global clustering coefficient}, sometimes also called {\em transitivity}, was first introduced by Barrat and Weigt in~\cite{barrat2000properties}. It is defined as 
\rejectforwg{3 times the number of triangles in the graph, divided by the number of connected triplets of vertices in the graph.}
\acceptforwg{\[ 
CC_g = 3 \times \frac{\text{\# triangles in the graph}}{\text{\# connected triplets of vertices in the graph}}. 
\]}
Another definition was given by Watts and Strogatz~\cite{watts1998collective} and is called the \textit{local clustering coefficient}. It is defined as the mean over all nodes of the graph of the local clustering of each node, that is the probability that two random neighbors of the node are also connected together. 
We use the global \cc in this paper. The \cc has also been defined for weighted graphs~\cite{saramaki2007generalizations,opsahl2009clustering}.
\end{conference}
\begin{journal}
The \textit{local clustering coefficient} of a node i, first introduced by Watts and Strogatz~\cite{watts1998collective}, is defined as the probability that two neighbors of i are also connected together. This probability can be computed as 
\[ 
CC(i) = \frac{\text{\# triangles with the node i}}{\text{\# connected triplets centered on i}}, 
\]
where (\# connected triplets centered on i) = $\binom{\text{deg(i)}}{2}$. From here can be defined for the whole graph a clustering coefficient as the mean of the local clustering coefficients over all the nodes of the graph: \[ CC_{g1} = \frac{1}{n} \sum \limits_{i \in V} CC(i) \]

Another definition was first introduced by Barrat and Weigt in~\cite{barrat2000properties}, and is called the \textit{global clustering coefficient}, or \textit{transitivity}. It is defined as 
\[ 
CC_g = 3 \times \frac{\text{\# triangles in the graph}}{\text{\# connected triplets of vertices in the graph}}. 
\]
We use the global \cc in this paper. The \cc has also been defined for weighted graphs~\cite{saramaki2007generalizations,opsahl2009clustering}.
%or directed graphs~\cite{fagiolo2007clustering}. \giovanni{And tcc, and ccc?} \thib{cf un peu en dessous.}
\end{journal}

%\begin{conference}
%\ms
%\end{conference}
\;

\rejectforwg{
\nitbf{Computations for random graphs.} 
%\giovanni{I would first introduce for random graphs, and then say it is higher for social network (donc, virer le debut de la phrase suivante au paragraphe d'apres).} \thib{(cf les deux prochains points violets pour faire ca.)}
%The \cc is usually much higher in social networks than in random models \thib{Remplacer avant par : The \cc had been theoretically computed in random models}, such as the Erd\"os–R\'enyi model~\cite{erds1960evolution} or the Barabasi-Albert model~\cite{barabasi1999emergence}. 
The \cc had been theoretically computed in random models, such as the Erd\"os–R\'enyi model~\cite{erds1960evolution} or the Barabasi-Albert model~\cite{barabasi1999emergence}. The values of the \cc for them respectively are $CC_{ER_{np}} = p$, with $p$ the (small) probability for two given nodes of the graph to be connected, and $CC_{BA} = \frac{M-1}{8} \frac{\log(n)^2}{n}$, with $M$ the degree of an arriving node and $n$ the number of nodes of the graph at the end of the process~\cite{bollobas2003mathematical}. }

%\begin{conference}
%\ms
%\end{conference}

\nitbf{Computations for social graphs.} The undirected clustering coefficient of some social networks has been provided in the literature. It has been computed on very large snapshots for Facebook \cite{ugander2011anatomy}, Microsoft Messenger~\cite{leskovec2008planetary}, Flickr, and YouTube~\cite{mislove2007measurement}. The local clustering coefficient has also been studied in the undirected mutual graph of Twitter~\cite{myers2014information}. 
% However, the values of the literature have been computed on small (and for most of them biased) samples of the graphs. In fact, the values largely differ between studies\footnotemark[3]. 
We can also cite the values 
%\footnotemark[3]\footnotetext[3]{\url{http://networkrepository.com/soc.php}} \giovanni{Footnote not needed since it can be recovered from [31]} 
given by the Network Repository project~\cite{rossi2015network}, providing a large comprehensive collection of network graph data available for which it lists some basic properties.
The undirected \cc is usually much higher in social networks than in random models. 

%\begin{conference}
%\ms
%\end{conference}
\;

\nitbf{Directed graphs.} All these studies only consider the undirected clustering coefficient, even for directed graphs like Twitter. 
Fagiolo introduced definitions of directed clustering coefficients, that we named \ccTTs and \ccCTs~\cite{fagiolo2007clustering}, but those definitions had never been computed and discussed on large datasets to our knowledge, as we do in this paper.
Moreover, we believe that these metrics are 
%{\em not very adequate to measure the clustering of directed graphs with interest links}. 
{\em not the most relevant ones for directed graphs with interest links.}
\;

\nitbf{Computing substructures.} Researchers studied methods to efficiently compute the number of triangles in a graph, as naive methods are computationally very expensive on large graphs. Two families of methods have been proposed: triangle exact counting or enumeration and estimations. \giovannivu{In what follows, say just one of them, and then "see also".} \thibvu{Ca me va de faire comme il dit... mais je te laisse faire cette partie ? :s} In the first family, the fastest algorithm is due to Alon, Yuster, and Zwick~\cite{alon1997finding} and runs in 
\begin{conference}
$O(m^{1.41})$, with m the number of edges. 
\end{conference}
\begin{journal}
$O(m^{\frac{2\omega}{\omega+1}})$, with m the number of edges and $\omega$ the best known exponent for the fast matrix multiplication. Its current value is 2.3728, due to an algorithm of~\cite{coppersmith1987matrix} improved by~\cite{le2014powers}, giving a complexity of $O(m^{1.41})$ for the AYZ algorithm. 
\end{journal}
However, methods using matrix multiplication cannot be used for large graphs because of their memory requirements. In practice, enumeration methods are often used, see e.g.,~\cite{latapy2008main,schank2005finding}. A large number of methods for approximate counting were proposed, see for example~\cite{kolountzakis2012efficient} and its references. The authors obtain a running time of $O(m + \frac{m^{3/2} \log n}{t\varepsilon^2})$   and a $(1 \pm \varepsilon)$ approximation. 
%, where n is the number of vertices and $m$ is the number of edges \thib{Enlever m vu que deja mis un peu plus haut ?}. 
%and $\Delta$ is the maximum number of triangles in which any single edge is contained. \fred{pb with this $\Delta$ which is not in the formula :-)}
Methods to count rectangles and butterfly structures in undirected bipartite networks were also proposed in~\cite{wang2014rectangle} and in~\cite{sanei2018butterfly}. 
In this paper, we propose an efficient enumeration algorithm to count the number of K22s and open K22s in a very large graph. We focused on the case in which only one adjacency can be stored, as this was our case for the \tss. To the best of our knowledge, we are the first to consider this setting.

\section{Computing Clustering Coefficients in Twitter}
\label{sec:computing}

%\acceptforwg{The Twitter snapshot (\tss in short), a directed network with 505 million nodes and 23 billion links that we use as a typical example of a directed social network with interest links, is described in annex~\ref{Annex:Annex_Twitter}.}
We computed the \icc and the triangle clustering coefficients on a directed Twitter snapshot (\tss in short) that we use as a typical example of a directed social network with interest links. 
We used two different methods: an exact count and an estimation using sampling techniques, either with a Monte Carlo algorithm or with a sampling of the graph.

\subsection{The Twitter Snapshot}
\label{Annex:Annex_Twitter}

In order to compute the different clustering coefficients of a real graph, the authors of \cite{gabielkov2012complete} gave us access to a snapshot of the graph of the followings of Twitter. The snapshot  was collected between March 2012 and July 2012. With $n=505$ million nodes and $m=24$ billion links collected, this graph is the largest directed social network graph available today, to the best of our knowledge. Each node of the graph represents an account of Twitter, and there is a link between two nodes $u$ and $v$, if the account $u$ follows the account $v$.  
All account IDs have been anonymized. 
%The graph is stored as an adjacency list containing, for each account, the list of its followers. 
%\fred{Taille exacte Adjacency list - should be discuss both way adjacency?}
The snapshot is a perfect case study as Twitter is a directed social network used both as a social and an information network~\cite{myers2014information,kwak2010twitter}. It allows to study directed/undirected social/interest clustering coefficients.
\\
\nitbf{Degree distributions of the \ts.}
\begin{figure}[t]
\centering
\includegraphics[scale=0.45]{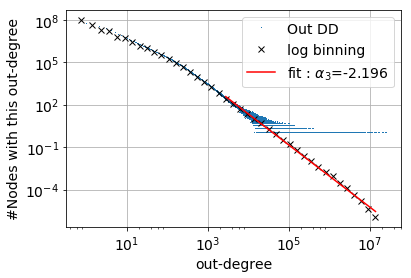}
\includegraphics[scale=0.45]{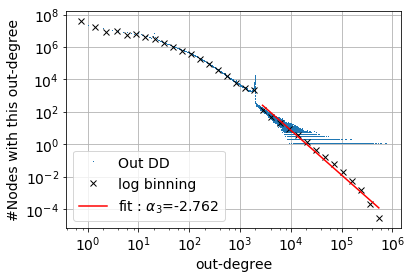}
\caption{In- (Top) and out-degree (Bottom) distributions of the \ts. The obtained distribution is given by the blue points; the black crosses represent the logarithmic binning of the distribution (a mean of a given amount of points on a logarithmic scale). The red straight line is the fit of the logarithmic binning; it has slopes of $-2.174$ and $-2.762$ for the in and out degree distribution.
\label{fig:twitter-power-law-fit}}
\end{figure}
We provide in Figure~\ref{fig:twitter-power-law-fit} the degree distributions of the \tss. 
We fitted their tails to power law distributions. 
We obtained $\Pmoins(i)=\Cmoins i^{-2.17}$ and $\Pplus(i)= \Cplus i^{-2.76}$, with $\Pmoins(i)$ (respectively $\Pplus(i)$) the probability that a node has in-degree (resp. out-degree) $i$. 
%We obtained $P_{in}(i)=A_{in} i^{-2.17}$ and $P_{out}(i)=A_{out} i^{-2.76}$, with $P_{in}(i)$ (respectively $P_{out}(i)$) the probability that a node has in-degree (resp. out-degree) $i$. 
In the following, we use the obtained values to compute the practical complexity of the algorithms.
\\
Other references of the literature have also provided a power law fit for both distributions, see e.g.,~\cite{myers2014information}. In this work, the authors obtained exponents of values 1.35 and 1.28. However, we believe that the authors did a fit on the complete distributions and not on their tails, leading to power law exponents below 2.
%which do not correspond to a preferential attachment model. 
This is why we preferred to only fit the tail. Another point of discussion would be to decide if the out-degree distribution really behaves as a power law. However, the best fit of the distributions is out of the scope of this paper. We just used the values provided by our fit as a possible model of the graph, but others exist.

%\vspace{-0.2cm}

%----------------------------------------------------------
% EXACT COUNT
%----------------------------------------------------------
%\newpage
\subsection{Exact Count}
\label{sec:exact-count}

We computed the exact numbers of K22s and open K22s in the Twitter Snapshot. Recall that we are discussing a dataset with hundreds of million nodes and billions of arcs. 
\begin{commentaires-imp}
\fred{should we put it? if yes, we should be coherent everywhere}
Since 21.3\% of the accounts of the \tss have a null in-degree and out-degree \cite{gabielkov2014studying}, if we remove those vertices, we are left to the study of a network with $n=396$ million nodes and $a=24$ billion arcs.
\end{commentaires-imp}
Results are reported in Table~\ref{tab:ccs} and discussed in Section~\ref{sec:results}. We also retrieved the number of directed and undirected triangles of \tss. We first discuss the complexity of algorithms for exact counting on very large graphs. We then present the algorithms we use and discuss the results. 

\acceptforwg{
\noindent
In the rest of this paper, we call \textit{top vertices} (resp. \textit{bottom vertices}) of a K22 the vertices which are destinations (resp. sources) of the K22 edges.  
%\giovanni{I would have made the opposite choice ;-)}. 
We call a {\em fork} a set of two edges of a K22 connected to the same vertex. We say that a {\em fork has top (or bottom) vertex $x$} if both edges are connected to $x$ and $x$ is a top (resp. bottom) vertex of the K22. The same terminology applies to open K22s.
}

%\ms
\;

\nitbf{Trivial algorithm.} The trivial algorithm would consider all quadruplets of vertices with 2 upper vertices. Then, for each quadruplet, it would check the existence of a K22 and of open K22s. There are $\binom 4 2\binom n 4$ such quadruplets. It thus gives a complexity of $O(n^4)$. This method can thus not be considered for the \tss as it would perform $\num{6.4e33}$ iterations.

%\ms
\;

\def\outn{\#occ(w)}
\def\internal{\#internalArcs}
\nitbf{Improved algorithm.} 
%\nitbf{Additional improvements.} 
\giovannivu{Not clear: this would work only to compute K22 but still you need to compute open K22; what are these? Additional refinements? Practical choices that control the constraints without changing the complexity?} 
\acceptforwg{The practical complexity can be greatly improved by only considering {\em connected quadruplets}, and by mutualizing the computations of the common neighbors of the in-neighbors of a vertex, as explained below.}
\rejectforwg{In fact, the practical complexity can be further improved by mutualizing the computations of the common neighbors of the in-neighbors of a vertex, as explained below. }
The pseudo-code is given in Algorithm~\ref{alg:compute-nk22s}.
\\
The algorithm's main loop iterates on the vertices of the graph. For each vertex $x$, we consider its in-neighborhood $N^-(x)$. We then compute how many times a vertex $w$ 
%(with $w\neq x$) 
(with $w < x$ to avoid counting a K22 twice)
appears in the out-neighborhoods of the vertices of $N^-(x)$. We denote it $\#occ(w)$. 
We use a hash table to store the value of $\#occ(w)$ in order to be able to do a single pass on each out-neighbor. 
%(corresponding to variable \outn{} in the pseudo-code) to store the results in order to be able to do a single pass on each out-neighbor. 
\\
For a vertex $w$, any pair of its $\#occ(w)$ in-neighbors common with $x$ forms a K22 with $x$ and $w$ as bottom vertices. There are hence $\binom k2$ K22s with $x$ and $w$ as bottom vertices. The number of K22s with $x$ as a top vertex is then
\[
\#K22(x) = \sum_{w|\#occ(w)\geq 2} \binom{\#occ(w)}2.
\]
%and the number of open K22s with the fork (vu,wu) is 
%\[\#openK22s(vu,wu) = d^+(v)-1+d^+(w) - 1 - \mathds 1_{w\in N^+(v)}-\mathds 1_{v \in N^+(w)},\]
The number of open K22s with $x$ as the top vertex is computed by noticing that,
for any pair of vertices $u$ and $v$ of $N^-(x)$, we have $d^+(u)-1+d^+(v) - 1 - \mathds 1_{v\in N^+(u)}-\mathds 1_{u \in N^+(v)}$ open K22s containing this fork $(ux,vx)$. We can count the number of open K22s with $x$ as a top vertex, $u$ as the bottom vertex of out-degree 2 (and thus another vertex $v$ as the bottom vertex of out-degree 1). A vertex $u\in N^-(x)$ is thus in $(d^+(u)-1\sum_{v\in N^-(x)\setminus \{u\}}1_{v\in N^+(u)})(d^-(x)-1)$ such open K22s. The only subtlety is that we count the number of  arcs, which are between two vertices of $N^-(x)$, during the loop on the out-neighborhoods of the vertices of $N^-(x)$. We note this number \internal{}. We then have: 
\[
\#openK22(x) = \left(\sum_{u\in N^-(x)} (d^+(u)-1)(d^-(x)-1)\right) - \internal. 
\]
%\begin{align*}
%\#openK22(x) = &\left(\sum_{u\in N^-(x)} (d^+(u)-1)(d^-(x)-1)\right)
%\\
%&- \internal. 
%\end{align*}
Lastly, the global number of K22s (resp. open K22s) in the digraph is just the sum of the number of K22s (resp. open K22s) with a vertex $x$ as a top vertex, as, since we only consider K22s formed with a vertex $w$ such that $x<w$, we only count each K22 once. 
% Last, the global number of K22s in the digraph is half the sum of the number of K22s with a vertex $x$ as a top vertex, as again, each K22 is computed twice. 
% \fred{On aurait pu eviter ca en considerant que les sommets $w>x$ dans la boucle  \thib{Let us notice that we could had avoid this by looping only on nodes $w>x$, which would had made the computation two times faster.}}

%\ms 

%\def\internal{{\tt internal\_arcs}}
\def\outn{\#occ}
\begin{algorithm}
\floatname{algorithm}{Algorithm}
\caption{Enumeration of K22s and open K22s\label{alg:compute-nk22s}}
\begin{algorithmic}[1]
\State $\rhd$ 
\State {\bf Input:} Digraph$(V,A)$
\State \outn=0 \Comment{hash table}
\For{$x \in V$}
\State $\internal \gets 0$ \Comment{We count the number of arcs internal to $N^-(x)$ as these arcs do not form open K22s}
\For{$v \in N^-(x)$}
\State \#openK22s $+= (d^+(v)-1)(d^-(x)-1)$
\For{$w \in N^+(v)\setminus\{x\}$}
\State \outn$[w] += 1$
\If{$w \in N^-(x)$} \Comment{We use a second hash table to test that.}
\State{$\internal += 1$}
\EndIf
\EndFor
\EndFor
\For{$w$ with \outn$[w] \geq 2$}
%\If{$\outn[w] \geq 2$}
\State \#k22$ += \binom{\outn[w]}2$
%\Else 
%\State \#k22$ += 1$
%\EndIf
\EndFor
\State \#openK22s $-= \internal$
\State $\outn \gets 0$ \Comment{Done with a double loop}
\EndFor
%\State \#K22 $\gets$ \#K22/2
\State \iccs $\gets \frac{4 \#K22}{\#open K22}$
\end{algorithmic}
\end{algorithm}

%\rejectforwg{Can be removed. }
\rejectforwg{
We further carried out several small optimizations in the main loop of the algorithm. For example, as the graph is stored as a table of lists of neighbors, we do not have access in $O(1)$ to the information that $u$ belongs to a neighborhood $N^-(x)$. We can get it in $O(\log_2(d-(x)))$ by doing a binary search.
%search by dichotomy.
We preferred to avoid this, and thus we use two hash tables:  
a first one storing the number of occurrences of a vertex $w$ in the out-neighborhoods of the vertices of $N^-(x)$, and a second one storing $N^-(x)$ to compute the internal arcs. 
}

\begin{commentaires-imp}
\fred{Should we put them in appendix?}
\begin{itemize}
\item As the graph is stored as a table of lists of neighbors, we do not have access in $O(1)$ to the information that $u$ belongs to a neighborhood $N^-(x)$. We can get it in $O(\log_2(d-(x)))$ by doing a search by dichotomy. We preferred to avoid this, and thus we use two hash tables:  
\begin{itemize}
\item A hash table to store the number of occurrences of a vertex $w$ in the out-neighborhoods of the vertices of $N^-(x)$. 
\item A hash table to store $N^-(x)$ to compute the internal arcs. 
\end{itemize}
\item We avoid some multiplications by: 
\begin{itemize}
\item computing $2 \cdot \#K22(x) = \outn[w]\break \cdot (\outn[w]-1)$. We then get at the end 2 times the number of K22s. We just have to multiply by 2 and divide by the \#openK22s to obtain \iccs. 
%\item computing $2 \cdot \#K22(x) = \outn[w]\break \cdot (\outn[w]-1)$. We then get at the end 4 times the number of K22s. We just have to divide by the \#openK22s to obtain \iccs. 
\item We avoid $d^-(x)$ multiplications of line 7 by doing it only once per vertex on line 14. 
\end{itemize}
\item We re-initialize the table \outn\ by using the same double loop on the out-neighborhoods of the vertices of $N^-(x)$ to avoid iterating on the table. 
\end{itemize}
\end{commentaires-imp}

%\ms
\;

\niem{Complexity of the used algorithm.}
The complexity thus is $m+\sum_u d^{+}(u)(d^{+}(u)-1)$. Indeed, each edge is only considered once as an in-arc and $d^{+}-1$ times as an out-arc. Note that, in the \ts, the sum of the squares of the degrees is equal to $8\cdot  10^{13}$. The order of the number of iterations needed to compute the number of K22s was thus massively decreased from the $\num{6.4e33}$ iterations of the trivial algorithm.

% \ms 

% \niem{Complexity on the \ts and on a preferential attachment model.} 
% In the \ts, the sum of the squares of the degrees is equal to $8\cdot  10^{13}$. 

%\ms
\;

\begin{shortconference}
%\niem{Complexity on a preferential attachment model.} %\thib{Macro $Cmoins$, $Pmoins$}
\niem{\thib{graph following a power-law degree distribution}}
A main characteristic of the real networks is their heavy-tailed degree distribution: most of them have a power-law degree distribution ~\cite{clauset2009power} ~\cite{newman2018networks}, with a power exponent between 2 and 3. It is known that the maximum degree of a power-law distribution $P(i)=C i^{-a}$ is in $O(n^{\frac{1}{a-1}})$ \cite{cohen2000resilience}. 
\end{shortconference}
\begin{conference}
%\niem{Complexity on a preferential attachment model.} %\thib{Macro $Cmoins$, $Pmoins$}
\niem{Complexity on graphs following a power-law degree distribution.}
\end{conference}
\niem{Complexity on graphs following a power-law degree distribution.}
The complexity of the algorithm on a graph built with preferential attachment can be computed as follows. 
We consider without loss of generality that the sum of the square of the degrees is minimum for the out-degrees (and not the in-degrees). 
The maximum degree is $\dpmax = O(n^{1/(\alpha^+-1)})$, with $\alpha^+$ the exponent of the out-degree power law distribution. 
%Thus, the sum of the squares of the degrees, when $2\leq \alpha^+ < 3$, is 
%$\sum_{v\in V} (d^+(v))^2 = C_{\alpha^+} n \sum_{i=1}^{\dpmax} \frac{i^2}{i^{\alpha^+}} \underset{n\to \infty}{\sim} C_{\alpha^+} n \int_{i=1}^{\dpmax} \frac 1{i^{\alpha^+-2}} = \left[ \frac {C_{\alpha^+} n}{(3-\alpha^+)i^{\alpha^+-3}}\right]_1^{\dpmax} = \frac{C_{\alpha^+} n}{(3-\alpha^+) {\dpmax}^{\alpha^+-3}}$\\
%$= \frac{C_{\alpha}}{(3-\alpha^+)} n^{1+\frac{3-\alpha^+}{\alpha^+-1}}$. 
%\giovanni{It looks like $C_\alpha + n$ instead of $C_{\alpha^+} n$...}
Thus, the sum of the squares of the degrees, when $2\leq \aplus < 3$, is 
$\sum_{v\in V} (d^+(v))^2 $ = $ \Cplus n \sum_{i=1}^{\dpmax} \frac{i^2}{i^{\aplus}}$ $\underset{n\to \infty}{\sim}$ $\Cplus n \int_{i=1}^{\dpmax} \frac 1{i^{\aplus-2}} $ = $ \left[ \frac {\Cplus n}{(3-\aplus)i^{\aplus-3}}\right]_1^{\dpmax}$ $\simeq$ $\frac{\Cplus n}{(3-\aplus) {\dpmax}^{\aplus-3}} $ = $
 \frac{\Cplus}{(3-\aplus)} n^{1+\frac{3-\aplus}{\aplus-1}}$, 
where $\Cplus=\frac{1}{\sum_{i \in \mathbb{N}^+} i^{\aplus}}$.
The complexity is thus in $O(m+n^{1+\frac{3-\aplus}{\aplus-1}})$. 
For preferential attachment graphs with exponents between 2 and 3, this gives a complexity between $O(m+n)$ and $O(n^2)$, to be compared to the one of the naive method $O(n^4)$. 
\begin{commentaires}
\thib{Discuter de l'AN}
\end{commentaires}
%For $\alpha=2.76$ (as fitted for the dout distribution of the \tss), we get $n^{1.14}$. As $n=\num{5.05e8}$, we obtain $\sim 8 \cdot 10^{9}$, which is lower than what we found in practice.
%of similar order as what found in practice.
\\
% The maximum degree is $\dpmax \underset{n\to \infty}{\sim} \frac{n^{1/(\alpha_o-1)}}{(\alpha_o-1)^{\alpha_o-1}}$. 
% Thus, the sum of the squares of the degrees, when $2\leq \alpha < 3$, is 
% $\sum_{v\in V} (d^+(v))^2 = n \sum_{i=1}^{\dpmax} \frac{i^2}{i^{\alpha}} \underset{n\to \infty}{\sim} \int \frac 1{i^{\alpha-2}} = \left[ \frac 1{(3-\alpha)i^{\alpha-3}}\right]_1^{\dpmax} = \frac{n}{(3-\alpha) {\dpmax}^{\alpha-3}}
% = \frac{n^{1+\frac{3-\alpha}{1-\alpha}}}{(3-\alpha)(\alpha-1)^{\alpha-1}}$. 
% For $\alpha=2.5$ (as fitted for the dout distribution of Twitter snapshot) \fred{it is written 2.5 here and 2.27 in the preliminaries}, we get $n^{4/3}$. As $n=4 \cdot 10^8$, we obtain $3 \cdot 10^{11}$, which is of similar order as what found in practice. \fred{to be adjusted... should we keep it}
% \thib{]  Je suis d'avis de l'enlever ; + ca permet de ne pas introduire la PL de Twitter $\rightarrow$ permet d'enlever aussi la partie qui en parle pour mettre dans papier de Guillaume.}
% \fred{to be discussed with thibaud}
\begin{conference}
Note that the number of undirected and directed triangles can be easily computed while counting the K22s. We do not present how here due to lack of space. 
\end{conference}
\begin{journal}

\ms

\nitbf{Counting the number of triangles.} The number of transitive triangles can easily be computed for free while counting the K22s. When iterating over the vertices of the TS and considering the vertex $x$ in Algorithm~\ref{alg:compute-nk22s}, the number {\tt internal\_arcs} of arcs between vertices of $N^-(x)$ corresponds to the number of transitive triangles for which $x$ is the top vertex.
The number of open transitive triangles with $x$ as the top vertex is simply $d^-(x)\cdot d^+(x)$. The total number of open transitive triangles is then just the sum of this quantity over all $x$. The number of cyclic triangles for $x$ can also be easily computed by counting the number of arcs from  $N^+(x)$ to $N^-(x)$. Each cyclic triangle is counted three times. The number of open cyclic triangles is the same as the number of transitive triangles. We can compute the number of undirected triangles with similar methods (either on the full (but undirected) graph or on the mutual graph). 

\begin{commentaires}
Complexity of the triangle counting could be computed precisely. 
\end{commentaires}
% The number of open triangles can be easily be computed using the indegree and outdegree distributions. The number of open transitive triangles is just the sum over all users of the product between their indegrees and outdegrees.

%\ms

\noindent
%nitbf{Discussion on the algorithm choice.} 
Note that the fastest methods to compute triangles in graphs have a complexity of $O(m^{1.41})$, where $m$ is the number of edges ~\cite{alon1997finding}. These methods rely on fast matrix multiplications and cannot be applied for large graphs as they need to have the full matrix in memory. Moreover, our algorithms would be faster in practice for large complex networks as they are sparse graphs. The average indegree (or outdegree) has a low value of 45.6 \cite{gabielkov2014studying} in Twitter. The complexity of the matrix methods would be of the order of $3.2 \cdot 10^{14}$ for the \tss as 
$m=2.3 \cdot 10^{10}$. This is higher than the practical complexity of computing the exact number of K22s (which is itself higher than the complexity of computing triangles). 
We discuss the obtained results with the exact count in Section~\ref{sec:results}. 
%\giovanni{In general I would separate this "results" section from the algorithm, this is a raw contribution, what do we learn on Twitter.}

\begin{commentaires-imp}
\fred{be careful $O(constant)$}
\end{commentaires-imp}

%\begin{commentaires}
%\fred{vérifier que ça répond mieux au commentaire de thibaud ci-dessous}
%\thib{(comment c'est possible du coup ? :o)}
%\end{commentaires}

% In the case of a sparse graph, which is the case of a social network. The average degree is 46 and can be considered constant with the number of vertices, see Figure~\ref{fig:avg_deg_evolution}. \fred{add the graph from Maksym}

% \begin{figure}
% \centering
% \includegraphics[width=\columnwidth]{figures/avg_deg_evolution}
% \caption{Evolution of the average degree of users in Twitter across time.}
% \label{fig:avg_deg_evolution}
% \end{figure}

\end{journal}

%\vspace{-0.2cm}

%----------------------------------------------------------
% MONTE CARLO
%----------------------------------------------------------
%\newpage
\subsection{Approximate Counts}
\label{sec:monte-carlo}

As discussed later in Section~\ref{sec:results}, the exact count of the number of K22s and open K22s in Twitter implies massive computations. 
This number can be estimated using Monte Carlo Method and/or computations on a sample of the graph. We discuss both methods below. 
One of our goals was to see how good computations made in the literature using smaller Twitter snapshots were.

%\begin{conference}
%\ms
%\end{conference}

\subsubsection{Exact \iccs on Twitter Samples.} 

\begin{figure*}
%\vspace{-.4cm}
%\includegraphics[width=\columnwidth]{figures/monte-carlo/samplings-9-datasets}
%\includegraphics[width=\columnwidth]{figures/monte-carlo/samplings-9-datasets-diff}
\includegraphics[width=.32\linewidth]{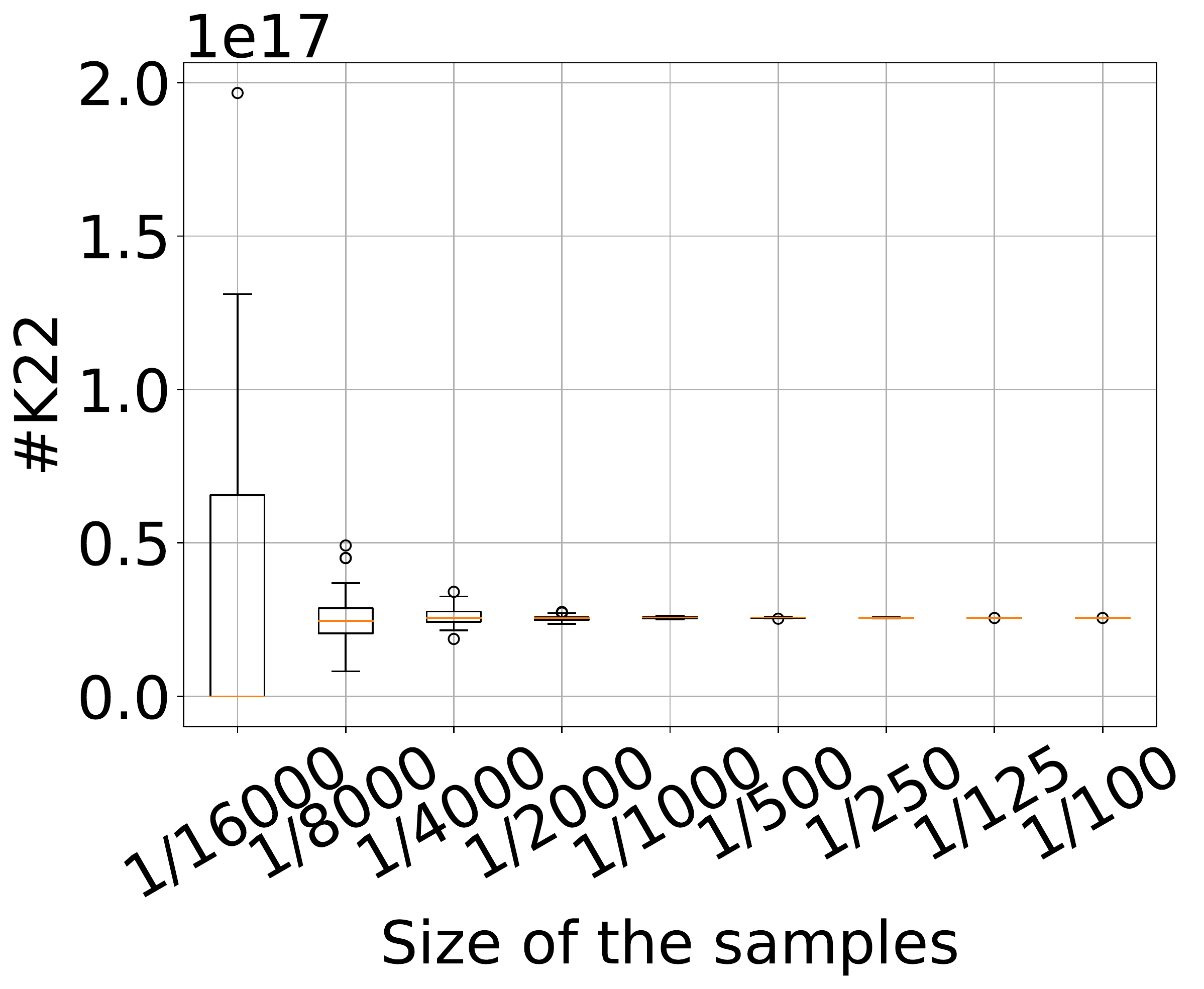}
\hfill
\includegraphics[width=.32\linewidth]{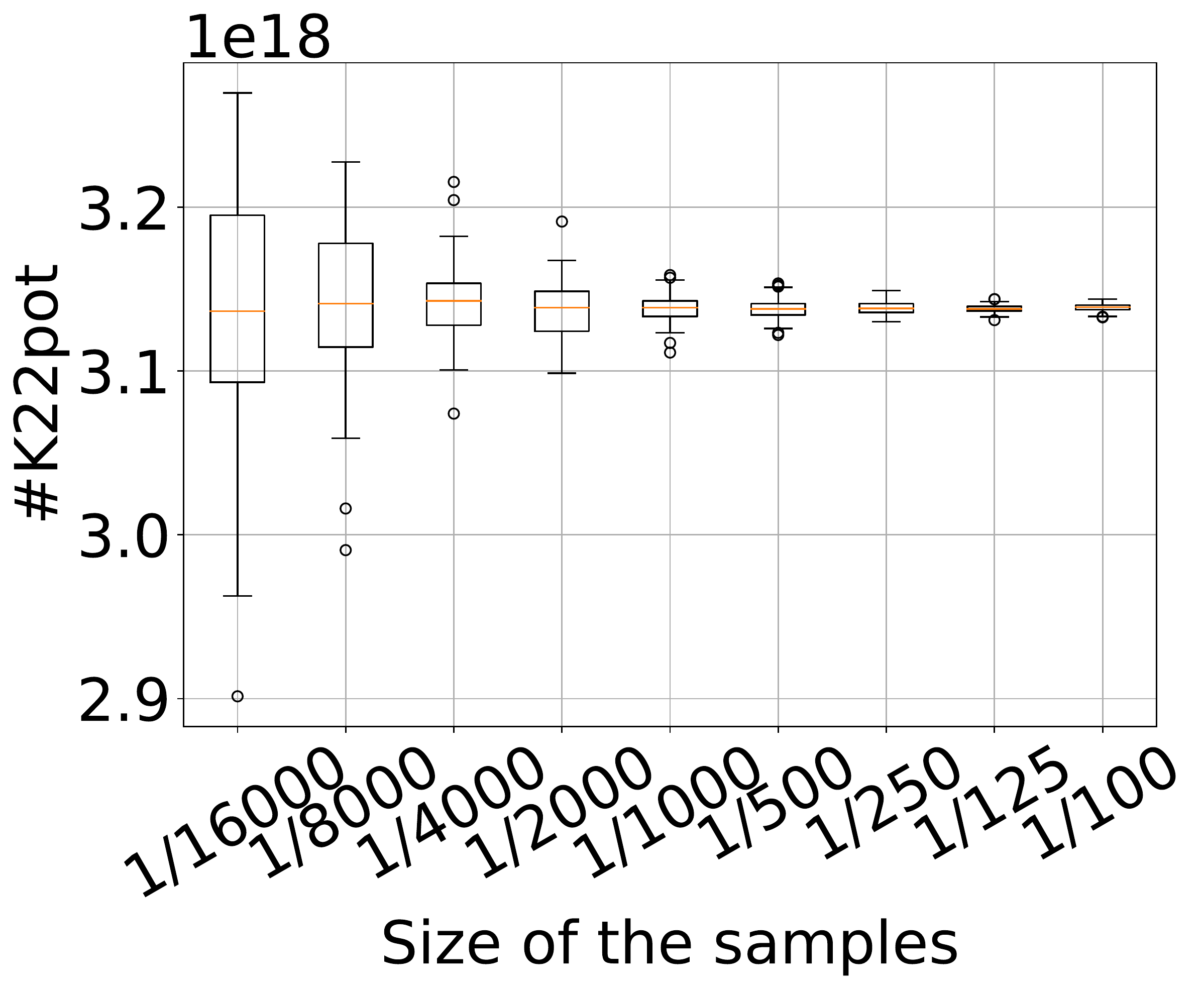}
\hfill
\includegraphics[width=.32\linewidth]{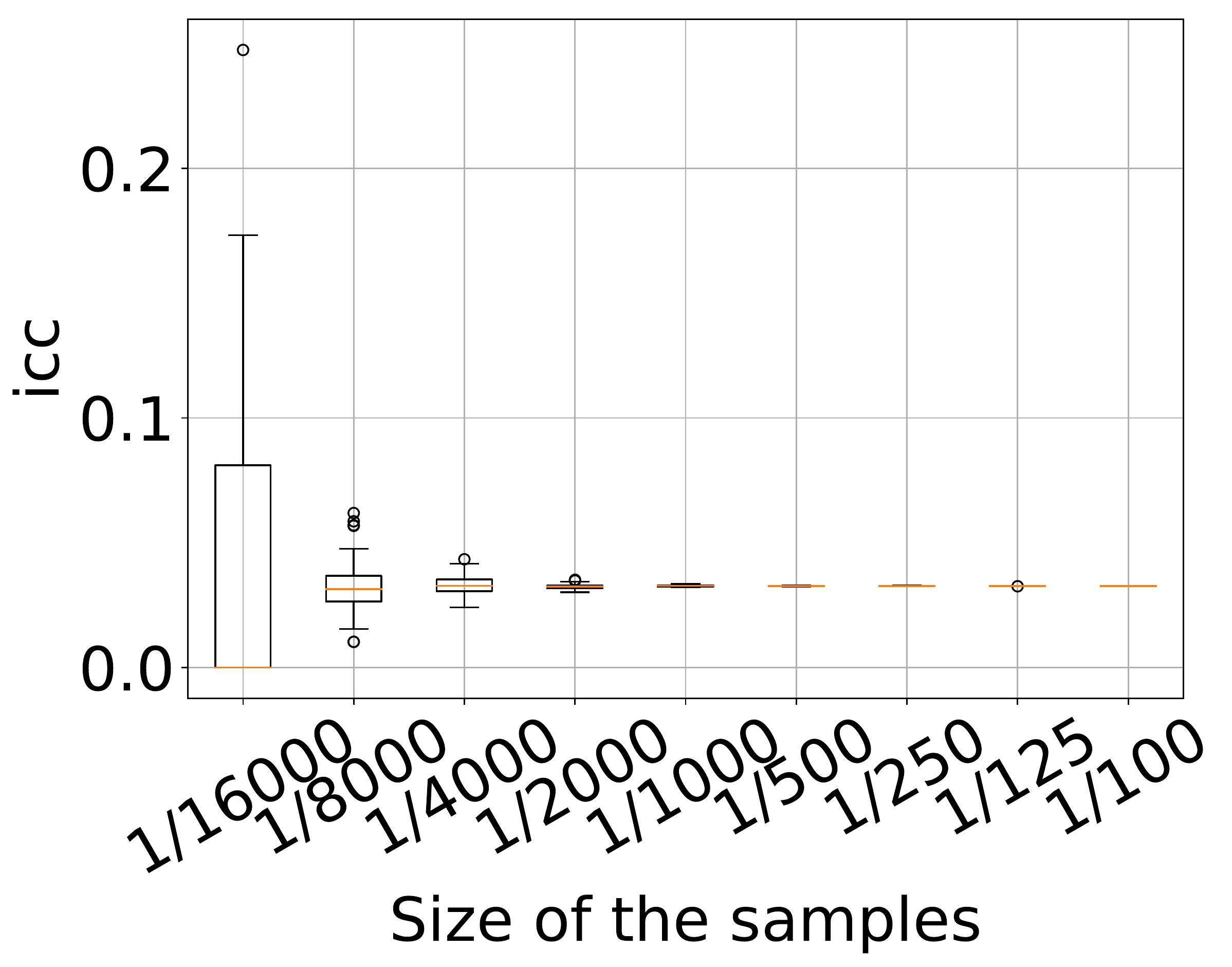}
\caption{Estimation of the K22s (Top), open K22s (Middle) and \icc (Bottom) for different sample sizes. \label{fig:samples}}
%\vspace{-.4cm}
\end{figure*}

We built samples of the \tss to estimate the \icc. Several choices can be made to build the samples. To avoid missing nodes of high degrees (which would lead to a high variance), we sampled the arcs (and not the nodes). Given a sampling probability $p$, we keep an arc in the sample with probability $p$. We generated samples of different sizes corresponding to sampling probabilities from $p=1/100$ to $p=1/16000$.

\begin{commentaires-imp}
\fred{the following is too detailed}
\end{commentaires-imp}

 \niem{Estimator of the number of K22 and open K22s.} Let us call $\mathcal A$ the set of occurrences of a specific pattern (in our case, either a K22 or an open K22). The number of occurrences of the pattern in a sample, $X$, is given by
$X = \sum_{A\in \mathcal A}  X_{A}$, 
where $X_A$ is the random variable which is equal to $1$ if all the arcs of pattern $A$ are selected in the sample and $0$ otherwise. 
\\
%\noindent
If we note $l$ the number of arcs of the pattern (4 for a K22 and 3 for an open K22), we have that $\mathcal P[X_A=1]=p^l$. By linearity of the expectation, we get
$E[X] = p^l |\mathcal A|.$
\begin{commentaires}
\[ E[X] = p^l |\mathcal A| \numberthis\label{EX}\]
\end{commentaires}
% if we call A a given pattern of the complete graph (in our case a closed or open K22), $X_{sub}$ the number of occurrences of this pattern in the subgraph, and $X_{sub,A}$ the random variable which is equal to $1$ if all the arcs of the pattern are selected by the sample and $0$ otherwise, we have : 
% \[ X_{sub} = \sum_{A}  X_{sub,A} \]
% Knowing that $X_{sub,A}$ is a Bernoulli variable verifying $E[X_{sub,A}]=p^l$ (where $l$ is the number of arcs of the pattern), and using the linearity of the expectation, we get 
% \[ E[X_{sub}] = p^l X, 
% \numberthis
% \label{EX}\]
% where $X$ is the number of occurrences of pattern A in the total graph. Since $l=4$ for the closed K22 and $l=3$ for the open K22, we have $E[\iccs_{sub}] = p \cdot \iccs$. 
Thus, $Y=p^{-l} X$ is an unbiased estimator of $|\mathcal A|$. %What is its efficiency? 
\begin{conference}
%\noindent
We succeeded to give theoretical bounds on the variance of this estimator. The difficulty was that the random variables $X_A$ are not independent, i.e., two K22s can share a common link.  Otherwise, the variance would simply be $\mathbb V(X)= \sum_{A\in \mathcal A}\mathbb V[X_A] = |\mathcal A| p^l(1-p^l) \leq |\mathcal A| p^l$. However, we argued that (and we later verified that), in practice, most of the K22s and open K22s do not share any link. 
\rejectforwg{We do not present the theoretical bounds here due to lack of space, but we show the estimator's efficiency in practice.}
\acceptforwg{The theoretical bounds can be found in annex~\ref{Annex:edge_sample}. }
\end{conference}

\begin{journal}
\niem{Variance.} Note that the random variables $X_A$ are not independent, i.e., two K22s can share a common link.  Otherwise, the variance would simply be $\mathbb V(X)= \sum_{A\in \mathcal A}\mathbb V[X_A] = |\mathcal A| p^l(1-p^l) \leq |\mathcal A| p^l$. However, we can argue that (and we will verify that), in practice, most of the K22s and open K22s do not share any link. It can be used in the analysis as follows. 
\begin{align*}
\mathbb V[X]   &= E[X^2] - E[X]^2 = E[(\sum_{A\in \mathcal A}X_A)^2]  - E[X]^2 \\ 
&= \sum_{(A,B)\in \mathcal A}  E[X_A X_B]  - E[X]^2
\end{align*}
\begin{longversion}
\begin{align*}
\mathbb V[X]   &= E[X^2] - E[X]^2 \\
&= E[(\sum_{A\in \mathcal A}X_A)^2]  - E[X]^2 \\ 
&= \sum_{(A,B)\in \mathcal A}  E[X_A X_B]  - E[X]^2
\end{align*}
\end{longversion}
We now distinguish the couples of dependent patterns, which we note $\Delta = \{(A,B) \mid A \cap B \neq  \emptyset\}$, from the ones of independent ones, ${\bar \Delta} = \{(A,B)  \mid A \cap B = \emptyset\}$.  
\begin{align*}
V[X]&= \sum_{(A,B)\in{\bar \Delta}}  E[X_A X_B]  +   \sum_{(A,B)\in \Delta}  E[X_A X_B]  - E[X]^2
\end{align*}
When $A$ and $B$ are independent, we have 
\[
E[X_A X_B]=\mathbb E[X_A]\mathbb E[X_B]=p^{2l}.
\]
As $E[X]^2=p^{2l}|\mathcal A|^2$, we get
\begin{align*}
V[X] &= \sum_{(A,B)\in{\bar \Delta}}  E[X_A]E[X_B]  +   \sum_{(A,B)\in \Delta}  E[X_A X_B]  - E[X]^2 \\
&= \sum_{(A,B)\in \Delta} (E[X_A X_B] - p^{2l})
\end{align*}
Let us now distinguish different cases. We note $\Delta_i$ the set of couples of patterns sharing $1\leq i \leq l$ arcs. For a couple $(A,B)\in \Delta_i$, we have that $\mathbb E[X_A X_B]=p^{2l-i}$, giving that 
\[
\mathbb V[X] \leq \sum_{i=1}^l \sum_{(A,B)\in \Delta_i} (p^{2l-i} - p^{2l}).
\]
Since $p<1$, we get
\[
\mathbb V[X] \leq \sum_{i=1}^l p^{2l-i}\left|\Delta_i\right|. \]
% If we suppose the worst case, i.e., that patterns $A$ and $B$ share all their arcs, we have that $\mathbb E[X_A X_B]=p^l$, giving
% \[
% \mathbb V[X] \leq \sum_{(A,B)\in \Delta} p^{l} - p^{2l}
% \]
% When $p$ is small, we get
% \[
% \mathbb V[X] \leq p^{l} \left|\Delta\right|. 
% \]
Note that, when all patterns are independent, $\left|\Delta \right|=|\Delta_l|=|\mathcal A|$ (couples $(A,A)\in \mathcal A$), giving back the variance of the independent case, $p^l|\mathcal A|$. \giovannivu{Do not understand the previous sentence.}
Chebycheff's inequality tells us that: 
\[ Prob [ |Y - \mu | \geq k \sigma ] \leq \frac{1}{k^2},\]
where $\mu$ is the expectation and $\sigma$ is the standard deviation of $X$. In our case, if we want an accuracy of $\varepsilon$ with a probability $q$, we should have $\frac{1}{k^2} \leq 1-q$ and $k \sigma \leq \varepsilon p^l |\mathcal A|$, 
which can be rewritten as: 
\[ 
\frac{k^2}{\epsilon^2} \sum_{i=1}^l p^{2l-i}\frac{|\Delta_i|}{|\mathcal A|^2} \leq p^{2l}. 
\]
\giovannivu{Not clear how the previous result is used.} \thibvu{On l'utilise dans numerical application... besoin d'expliquer dès maintenant en plus ?}
Lastly, to estimate the \iccs, we use as an estimator 
\[
Z = \frac{4Y}{Y_0}, 
\]
with $Y$ and $Y_o$ the estimators of the number of K22s and open K22s, respectively.  
As $\lim_{n\to \infty} Y = \#K22s$ and $\lim_{n\to \infty} Y = \#openK22s$, we have that 
$\lim_{n\to \infty} Z = \iccs$. 
For the precision, if $Y$ and $Y_o$ have an accuracy of $\varepsilon$ and $\varepsilon_o$ respectively, then with a probability $q=0.99$, $Z$ has at least an accuracy of $\frac{1+\varepsilon}{1-\varepsilon_o}\underset{\varepsilon \to 0}{\sim} 1+\varepsilon+\varepsilon_o$ with a probability $q^2\approx 0.98$.
\giovannivu{I have probably misunderstood what $\Delta_l$ is because I am not able to follow the reasoning and calculations here.}

\ms 

\niem{Numerical application.} We now consider the K22s of the \tss. Note that we know that $\frac{|\Delta_4|}{|\mathcal A|^2}=1/\#K22s=\num{3.8e-17}$. We also can notice that $|\Delta_3=\Delta_4|$. 
%As there are \num{2.6e16} K22s and 23 billion edges, an edge is shared by \num{1.1e8} K22s  in average. 
In the \tss, an edge is shared by $\frac{\#K22s}m$ K22s on average, with $m$ the number of links of the \tss. Thus, the average number of K22s sharing at least an edge with a K22 is between $\frac{\#K22s}m$ and $4\cdot \frac{\#K22s}m$. It gives $\frac 1{m}|\mathcal A|^2 \leq \Delta_1+\Delta_2+\Delta_3+\Delta_4 \leq \frac 4{m}|\mathcal A|^2$.
%Thus, if we consider a K22, 
% it shares an edge with in average 
% the probability that a given edge is shared by another given K22 is $\num{10e7}/\num{10e17}$. If we consider that in the \tss the correlation is 100 times stronger. 
The number of overlapping K22s with $i$ arcs is a non-increasing function of $i$. To make a numerical evaluation, we suppose that most overlapping K22s share one edge and not 2 edges in the \tss. We set that $|\Delta_1|=\frac 1m = \num{4.3e-8}|\mathcal A|^2$, and  $|\Delta_2|=10^{-16} |\mathcal A|^2$. Now, if we want a precision of $\varepsilon=0.1$ with a probability 0.99 (that is $k=10$), we need to take a sampling probability $p$ such that
\[
%p^8 \geq \frac{10^2}{10^{-4}}(p^{7}\num{4.3e-8} + p^{6}\num{e-16} + p^{5}\num{3.8e-17} + p^{4}\num{3.8e-17}).
p^8 \geq \frac{10^2}{10^{-4}}(p^{7}\num{4.3e-8} + p^{6}\num{e-16} + p^{5}\num{3.8e-17} + p^{4}\num{3.8e-17}).
\]
That is $p\geq \num{2.5e-4}$. 
%\begin{commentaires-imp}
%\thib{Résolution numerique, préciser ?}
%\end{commentaires-imp}
%epsilon 0.01 : 0.0435
%epsilon 0.1: 0.0002493
% where $\mu$ is the expectation and $\sigma=\sqrt{Var[X]}$ is the standard deviation. In our case, if we want an accuracy of $\varepsilon$ with a probability $q$, we should have $\frac{1}{k^2} \leq 1-q$ and $k \sigma \leq k p^{\frac{l}{2}} \sqrt{\Delta} \leq \varepsilon p^l |\mathcal A|$, 
% which can be rewritten as : 
% \[ \frac{k^2}{\epsilon^2} \frac{|\Delta|}{|\mathcal A|^2} \leq p^{l} \]
% \niem{Numerical application.} If we consider most patterns  are disjoint and set $\frac{|\Delta|}{|\mathcal A|^2}=10^{-8}$ (we will see in the following that, in fact, the fraction is lower in the Twitter Snapshot), if we want a precision of $\varepsilon=0.1$ with a probability 0.99 (that is $k=10$), we need to take a sampling probability $p$ such that
% \[
% p^4 \geq \frac{10^2}{10^{-2}}10^{-8}.
% \]
% That is $p\geq 1/10$. 
Thus, under these hypotheses, a sample with sampling probability 1/2500 and larger, e.g., our 1/2000 sample, allows to estimate the number of K22s with a precision of 10\%. The number of open K22s is larger and thus, the precision is better.
It gives a precision of at least  $\frac{1+1/100}{1-1/100}=0.20$ for the estimation of \iccs. In practice, the Chebysheff inequality and our hypothesis are pessimistic as shown below. 
\end{journal}

%\begin{conference}
%\ms
%\end{conference}
\nitbf{Results.} 
We present in Figure~\ref{fig:samples} the results of the algorithm for different sample sizes, corresponding to sampling probabilities from $p=1/100$ to p=$1/16,000$. For each sample size, we generated 30 samples. The distribution over the samples of the \icc, K22s and open K22s are provided by a boxplot for each value of $p$.
%of the estimated numbers of open and closed K22s and of the \icc is provided by a boxplot for each value of $p$. 
Note that a K22 of the \tss appears in a sample with a probability of only $p^4$, and of $p^3$ for an open K22. The clustering coefficient of a sample is thus an estimate of $p\cdot \iccs$. \giovannivu{Is this coherent with $Z = \frac{4Y}{Y_0}$?}
\\
We observe that the clustering coefficient is well estimated using any sample for a sampling probability of $1/1000$ or larger. Indeed, for this range of probabilities, the distribution over all samples is very concentrated and around the exact value of the \iccs. Note that, for $p=1/1000$, a K22 is present in the sample with a probability of only $10^{-12}$. The expectation of the number of nodes with an edge is only 23 million nodes (over 500 million) and the number of edges also around 23 million. Thus, a small sample (5\% of the nodes and 0.1\% of edges) allows to do an efficient estimation of the \iccs. 
\\
For smaller values of $p$, the variance increases. The median estimates well the \iccs for a range of $p$ between $1/8000$ and $1/1000$, but samples of these sizes may have error of 100\% of the value. Lastly, for $p=1/16000$, only the number of open K22s (and not the K22s or the \iccs) is approximated by the median.  
\\
In conclusion, a sample with sampling probability 1/1000 is enough to efficiently estimate the \icc, with a computation time of around 1 minute (instead of days for the whole \tss) on a machine of the cluster.

% \begin{itemize}
% \item Note that, for $p=1/1000$, a K22 is present with a probability of only $10^-{12}$. The expectation of the number of nodes with an edge is only ... 
% \item discuss variation over samples
% \item discuss variation over sampling probability
% \item discuss execution times
% \end{itemize}

% \fred{would be nice to do and discuss. Other sizes of samples. 1/10000? size of the graph in terms of number of vertices and arcs}

% \begin{figure}
% \vspace{-.4cm}
% %\includegraphics[width=\columnwidth]{figures/monte-carlo/samplings-9-datasets}
% %\includegraphics[width=\columnwidth]{figures/monte-carlo/samplings-9-datasets-diff}
% \includegraphics[width=\columnwidth]{Boxplot_Samples_K22.pdf}
% \includegraphics[width=\columnwidth]{Boxplot_Samples_K22pot.pdf}
% \includegraphics[width=\columnwidth]{Boxplot_Samples_CCi.pdf}
% \caption{Estimation of the K22s (Top), open K22s (Middle) and \icc (Bottom) for different sample sizes. \label{fig:samples}}
% \vspace{-.7cm}
% \end{figure}

%----------------------------------------------------------
% MONTE CARLO - 2
%----------------------------------------------------------
%\newpage
\subsubsection{Monte Carlo Method.} 

\begin{figure}[t]
%\vspace{-.4cm}
%\includegraphics[width=\columnwidth]{figures/monte-carlo-one-run-10M}
\begin{center}
    \includegraphics[width=1.\linewidth]{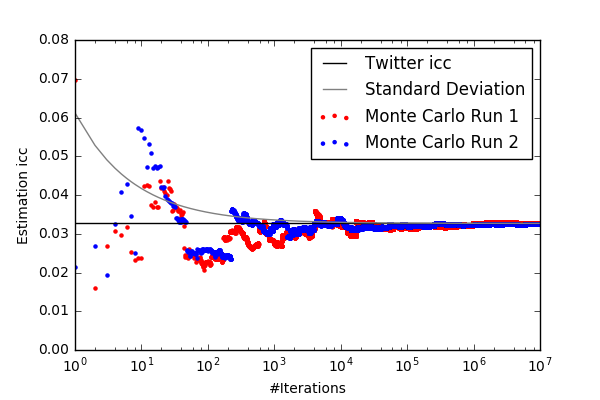}
	\caption{Estimation of the clustering coefficient with Monte Carlo Method. \label{fig:monte-carlo}}
\end{center}
%\vspace{-.7cm}
\end{figure}

\begin{journal}
After a short reminder of the precision of the Monte Carlo Method, we first quickly discuss the case of triangles to show the particularity of estimating the \icc. The difficulty here is that the probability to observe a (closed or open) K22 or a triangle is very small.  
In the case of triangles, this difficulty can be easily circumvented by knowing the node degrees. This allows to select an open triangle uniformly at random. In the case of K22s, this information is not sufficient to select an open K22 uniformly at random. In fact, achieving this goal is very costly, but we present a method in which, by picking only forks (as we do for triangles), we can compute the \icc.
\end{journal}
\begin{conference}
\rejectforwg{We first quickly discuss the case of triangles to show the particularity of estimating the \icc. }
The difficulty to estimate the clustering coefficients using Monte Carlo Method is that the probability to observe a (closed or open) K22 or a triangle is very small. 
In the case of triangles, this difficulty can be easily circumvented by knowing the node degrees. This allows to select an open triangle uniformly at random. In the case of K22s, this information is not sufficient to select an open K22 uniformly at random. In fact, achieving this goal is very costly.  \rejectforwg{We thus present a {\em new method in which, by picking only forks} (as we do for triangles), we can compute the \icc.}
\acceptforwg{We thus introduce a {\em new method in which, by picking only forks} (as we do for triangles), we can compute the \icc. This new method, as much as its theoretical justification, is developed in annex~\ref{Annex:MC}. 
The idea is to select a vertex $v$ as a root according to the square of its in-degree (as in the case of triangles), but without knowing its number of open K22s (first step). 
We then select two arcs $u_1v$ and $u_2v$ uniformly at random (second step). We then compute the number of K22s and open K22s with the selected fork $(u_1v,u_2v)$ (third step). 
We present here the results of the experiments.
}

%\ms

\end{conference}

\begin{journal}

\ms

\nitbf{Preliminary: Precision of Monte Carlo Method.}
\noindent{\em Precision of the estimation and number of iterations.} Each trial is a Bernoulli variable with probability $p$. We use as an estimate $Y$, the mean of the random sample. Its expectation is $p$ and its standard deviation is $\frac{\sqrt{p(1-p)}}{\sqrt n}$.
Due to the central limit theorem, we get that, when $n$ is large, 
\[ Prob\left[|Y-p| \leq Z_{\alpha/2} \frac{\sqrt{p(1-p)}}{\sqrt n}\right] = \alpha,\] 
with $Z_{\alpha/2}$ the value giving the $\alpha$ confidence interval 
a standard normal distribution. 
To get with probability $\alpha$ an accuracy of $\varepsilon$ of the empirical mean $p$ (which is not known), we should have $Z_{\alpha/2} \frac{\sqrt{p(1-p)}}{\sqrt n}\leq \varepsilon p$. That is $n \geq \frac{Z_{\alpha/2}^2(1-p)}{p\varepsilon^2}$. If we take $n \geq \frac{Z_{\alpha/2}^2}{p\varepsilon^2}$, we have the wanted precision (and we are not doing many more iterations when $p$ is small). For example, to get an accuracy of $99\%$ ($\varepsilon=0.01$), with probability $\alpha=0.99$, we should have a number of iterations such that $n \geq \frac{75,625}{p}$. 
%$np \geq Z_{\alpha/2}(1-p)/\varepsilon^2$. $np$ is the expected number of successes of the Bernoulli variables. Thus, we can stop after $Z_{\alpha/2}(1-p)/\varepsilon^2$ successes. For example, to get an accuracy of $99\%$ ($\varepsilon=0.01$), with probability $\alpha=0.99$, we can stop after 27,500 successes (corresponding to fewer than 850,000 iterations to estimate the K22 clustering coefficient of TS). \fred{It may be better not to use this notion of number of successes.}

\ms 

%\noindent\fred{The concentration given by Chebycheff's inequality has to be removed, but kept in comment as an exercise for Thibaud ;) or me ;)}
% Chebycheff's inequality tells us that: 
% \[ Prob [ |Y - \mu | \geq k \sigma ] \leq \frac{1}{k^2}.\]
% That is 
% \[ Prob [ |Y - p | \geq k  \frac{\sqrt{p(1-p)}}{\sqrt n}] \leq \frac{1}{k^2} \] 
% To get with probability $q$ an accuracy of $\varepsilon$ of the empirical mean, we should have $\frac 1{k^2}\leq 1-q$ and  $\frac{k\sqrt{p(1-p)}}{\sqrt n} < \varepsilon p$. That is $k \geq 1/\sqrt{1-q}$ and $n\geq \frac{k^2(1-p)}{p\varepsilon^2}$. 
% If we take $n\geq \frac{k^2}{p\varepsilon^2}$, we have the wanted precision (and we are not doing many more iterations when $p$ is small). 
% When $p$ is not known, we also have $np\geq \frac{k^2}{\varepsilon^2}$. Knowing that np is the number of times Thus, we can stop when $Y\geq \frac{k^2}{\varepsilon^2}$. 
% So, to get an accuracy of $10\%$ ($\varepsilon=0.1$), with probability $q=0.99$ (giving $k=10$), we can stop after 
% 10,000 successes (corresponding to 3,067,500 iterations for a the K22 clustering coefficient). 
% %should do $n=300,000$ iterations. Note that if 
% \fred{discuss gaussian approximation??}
\end{journal}
% SHOULD NOT BE SHORT CONFERENCE

\nitbf{Approximating the number of undirected triangles.} 
A first direct method would be to select three vertices uniformly at random and check if they form a triangle and open triangles. The problem with this method is that the probability to form a triangle in Twitter is the number of triangles divided by the number of triplet of nodes, i.e., $\frac{\num{6.23e11}}{(\num{5e8})^3}=\num{5e-15}$.
Thus the number of needed iterations would be astronomic, $\num{5.5e19}$ for an accuracy of $1\%$, with probability $\alpha=0.99$. We thus have to use methods selecting open triangles directly. 

 To estimate the \ccUT{}, we need to select open (undirected) triangles uniformly at random. We then test if the selected triangle is closed or not (which is the case with probability \ccUTs). The number of open triangles rooted at vertex $v$ is equal to $\frac{d(v)d(v)-1}2$. 
We can thus perform the sampling by picking a vertex $v$ with probability $\binom{d(v)}{2}/ \sum_{v \in V} \binom{d(v)}{2}$ and then select two random edges adjacent to $v$.

\ms 

\nitbf{Directed triangles.} The method is the same in the case of directed triangles. We select an open triangle uniformly at random. The number of open triangles rooted 
at a vertex v is $d^-(u)d^+(u)$. We thus select a node $u$ with probability $d^-(u)d^+(u)/ \sum_{v \in V} d^-(v)d^+(v)$. We then select uniformly at random an incoming arc and an outgoing arc. Lastly, we check if the triangle is closed (which is the case with a probability equal to \ccTTs and to \ccCTs respectively for transitive and cyclic triangles). 
%\fred{think about multiple count of a single triangle???}

\noindent{\em Precision of the estimation and number of iterations.} Each trial is a Bernoulli variable with a  probability $p=\ccTTs=0.019$. To get an accuracy of $1\%$, with probability $0.99$, we should thus do $n=\num{4e6}$ iterations.

\ms 

\nitbf{\Icc.} For triangles, we were able to select uniformly at random open triangles using the node degrees. In the case of K22s, node degrees is not sufficient to select an open K22 uniformly at random. To do so, it would be necessary to compute the number of open K22s with $u$ as a root. This pre-processing is very costly: for each node, we should consider its in-neighbors, sum their out-degrees, and compute the number of internal edges. It would be almost as costly as doing an exact count of the number of K22s. 

Another method is to select a vertex $v$ as a root according to the square of its in-degree (as in the case of triangles), but without knowing its number of open K22s (first step). 
% END OF WHAT SHOULD NOT BE IN THE SHORT CONFERENCE VERSION
We then select two arcs $u_1v$ and $u_2v$ uniformly at random (second step). We then compute the number of K22s and open K22s with the selected fork $(u_1v,u_2v)$ (third step). 

For the first step, the algorithm needs a list of the node in-degrees of the \tss, which would have been computed in a preliminary step. For the second one, it then uses the in-adjacency of $v$. For the third step, the out-adjacency of $u_1$ and $u_2$ are necessary for the computations.

We then use the estimators introduced below. We first define  
\begin{shortconference}
$X = \#K22s(u_1v,u_2v)$ and $X_o = \#openK22s(u_1v,u_2v).$ 
\end{shortconference}
%\begin{conference}
\[
X = \#K22s(u_1v,u_2v)
\quad \mbox{ and } \quad 
X_o = \#openK22s(u_1v,u_2v). 
\]
%\end{conference}
% \[
% X = \#K22s(u_1v,u_2v)
% \]
% and 
% \[
% X_o = \#openK22s(u_1v,u_2v). 
% \]
We have 
\[
\mathbb E[X] = \sum_{forks} \#K22s(fork) \mathbb P(fork). 
\]
As each fork is chosen uniformly at random and as a K22 has two forks, we get 
\[
\mathbb E[X] = \sum_{forks} \#K22s(fork) \frac 1{\#forks} = \frac{2 \#K22s}{\#forks}.  
\]
Similarly, 
\[
\mathbb E[X_o] = \frac{\#openK22s}{\#forks}.  
\]
We may thus define two efficient unbiased estimates for \#K22s and \#openK22s: 
\[
Y = \frac{\#forks}{2n} \sum_{i=1}^n X_i.
\quad \mbox{ and } \quad 
Y_o = \frac{\#forks}n \sum_{i=1}^n {X_o}_i.
\]
% \[
% Y = \frac{\#forks}{2n} \sum_{i=1}^n X_i.
% \]
% and 
% \[
% Y_o = \frac{\#forks}n \sum_{i=1}^n {X_o}_i.
% \]
We have $\mathbb E[Y] = \#K22s$ and $\mathbb E[Y_o] = \#openK22s$. 
The number of forks with a vertex $v$ as a root is given by $\binom{d^-(v)}{2}$. The total number of forks in the \tss is thus $\sum_{v\in V} \binom{d^-(v)}{2}$. \giovannivu{Why this observatin here? (les deux dernières phrases)}
\begin{commentaires-imp}
Note that $\sigma(X) = \sigma(\#K22s per fork)$ and thus $\sigma(Y) = \frac{\sigma(\#K22s per fork)}{\sqrt n}$. 
Similarly, $\sigma(Y_o) = \frac{\sigma(\#openK22s per fork)}{\sqrt n}$. \fred{unfortunatly, we do not have this number, we have something a little bit similar, the number of K22s and open K22s per vertices along with their degrees...}
\fred{BUT WE HAVE THE ONE FROM MONTE CARLO EXP}
\end{commentaires-imp}
Lastly, as we are interested by the \icc, we define
\[
Z = \frac {4Y}{Y_0}.
\]
As $\lim_{n\to \infty} Y = \#K22s$ and $\lim_{n\to \infty} Y = \#openK22s$, we have that 
$\lim_{n\to \infty} Z = \iccs$. 

\begin{commentaires-imp}
\fred{discuss precision}
Note that if the precision for the estimations of \#K22 and of \#openK22s are $\varepsilon_1$ and $\varepsilon_2$, since ($\frac{1+\varepsilon_1}{1-\varepsilon_2} \underset{\varepsilon \to 0}{\sim} 1+\varepsilon_1+\varepsilon_2$) the precision of $Z$ to estimate \iccs is $\varepsilon_1+\varepsilon_2$. 
\end{commentaires-imp}

\ms

\nitbf{Experiments.} We carried out two runs with 10 million iterations. It took about 2min30 for one run (\num{60,000} iterations per second). The value of 
\rejectforwg{$Z$}
\acceptforwg{the estimator of the \iccs}
for the two runs is plotted as a function of the number of iterations in Figure~\ref{fig:monte-carlo}. We first see that the estimator converges as expected to the value of the \iccs of \tss represented by a straight horizontal line (and which was computed exactly in the previous section). We also plotted the estimated standard deviation as a function of the number of iterations. To obtain it, we did one billion iterations. 
%We then estimated the standard deviation using the estimate $\sigma = \sqrt{\frac{1}{n-1}\sum_{i=1}^{10^9} (\frac{X_i}{{X_o}_i} - \iccs)^2}$. We then plot $\frac{\sigma}{\sqrt n}$.
We then estimated the standard deviation $\sigma$, and plotted $\frac{\sigma}{\sqrt n}$.
We see that large jumps or discontinuity happen, but only at the beginning. They correspond to the draw of a fork with a lot of K22s and open K22s corresponding to a user who does not have the same \iccs as the global network. 
\rejectforwg{Indeed, as shown later in Section~\ref{sec:exact-count}, nodes with very different values of \iccs exist (Figure~\ref{fig:Distrib_cc}).}
Then, the convergence is quick. After 300 iterations, the standard deviation is below 10\% and after 1000 iterations, we do not experience a value of the runs less precise than 10\%. 
\begin{commentaires}
.. \fred{maybe give how many iterations for an average mistake of 10\%, 1\%...} 
\end{commentaires}
\begin{commentaires-imp}
Note that we have seen in Section~\ref{sec:exact-count} in Figures~\ref{fig:by-nodes/cc-distribution} that nodes 
%with very different \iccs exist as well as nodes with a very large number of K22s and open K22s, see Figures~\ref{fig:by-nodes/k22s-distribution} and~\ref{fig:by-nodes/openk22s-distribution}.    
with very large numbers of K22s and open K22s, see Figures~\ref{fig:by-nodes/k22s-distribution}.  and~\ref{fig:by-nodes/openk22s-distribution}.   
\end{commentaires-imp}

\begin{table*}[t]
\begin{center}
    $
	\begin{array}{|l|ccc|}
	\hline
	& \# closed &\#open & cc\\
	\hline
	\iccs & 25,605,832,012,451,571 & 3,138,466,676,914,054,233 & 0.032634831\\
	& 2.6 10^{16} & 3.1 10^{18} & 3.3\%\\
	tcc & 2,469,018,039,988 & 129,023,573,841,024 &  0.019136178\\
	& 2.5 10^{12} & 1.3 10^{14} & 1.9\%\\
	ccc & 723,131,368,202 & 129,023,573,841,024 &  0.016813936 \\
	& 7.2 10^{11} & 1.3 10^{14} & 1.7\%\\
	ucc & 623,873,346,660 & 1,631,948,600,661,523 &  0.001146862\\
	 & 6.23 10^{11} & 1,63 10^{15} &  0.11\%\\
	mcc & 317,649,850,664 & 8,924,125,201,234 &  0.106783526\\
	& 3.2 10^{11} & 8.9 10^{12} & 10.7\%\\
	\hline
	\end{array}
	$
\caption{Clustering coefficients (exact and approximated count) in the \tss. \label{tab:ccs}}
\end{center}
%\vspace{-1.cm}
\end{table*}

% FROM LUC
% The computation of the sample graphs, starting from 1/2 to 1/32000 has been done using heavy parallelism on computers equipped with 16 to 48 cores and 192GB of RAM.

% Because of the RAM limitation, computing the entire graph had to be done using multiple computers in the NEF cluster at Inria. The problem was divided in a set of 5.000 subproblems, which were treated in parallel using a batch middleware. This asynchronous computation employed about 30 computers with similar hardware and took a total of 4 days.

% The search algorithms require the knowledge of both graphs direction. Unfortunately each adj table takes 100GB RAM. This means that no two ADJ table cal be loaded at same time. The counting algorithms then operate by iterating on the IN-adj, loading it page after page, keeping only 1 page in RAM. This leaves enough RAM to allow the loading of the entire OUT-adj, which is then randomly accessed.

% We made an extensive use of a rack of 16 Dell C6420 dual-Xeon 2.20GHz (20 cores), with 192GB RAM, all sharing an NFS Linux partition over Infiniband.

% R     14:38:58     Duration: 51h 19min 3s
% R     14:38:58     Cumulated computation time: 265h 38min 49s
% R     14:38:58     
% R     14:38:58      - nbK22=25605832012451817
% R     14:38:58      - nbK22pot=3138466676914054233
% R     14:38:58      - CK=0.03263483050599619
% R     14:38:58      - nbTransitiveTriangles=2469018039988
% R     14:38:58      - nbTransitiveTrianglesPot=64511786920512

\section{Results: Clustering coefficients in Twitter}
\label{sec:results}

To compute the number of K22s and open K22s, directed triangles, and undirected triangles in the \ts, we used a cluster with a rack of 16 Dell C6420 dual-Xeon 2.20GHz (20 cores), with \machineMemory RAM, all sharing an NFS Linux partition over Infiniband. It took 51 hours to compute the exact numbers of K22s and open K22s, corresponding to 265h of cumulative computation times on the cluster. We reported the results in Table~\ref{tab:ccs}. 

%\begin{conference}
%\ms
%\end{conference}

\nitbf{Number of K22s and triangles.} We see that the numbers of K22s and open K22s are huge, \num{2.6e16} and \num{3.1e18}, respectively. It has to be compared with the number of triangles which are several orders of magnitude smaller: e.g., \num{2.5e12} and \num{1.3e14} for transitive triangles. 
%\thib{Virer tout ce qu'il y a après : il n'y a plus rien en appendice !! Et du coup, reformuler un peu ce qu'il y a avant.} For the curiosity of the reader, the distributions of the number of closed and open K22s per user are provided in Figure~\ref{fig:k22-distributions} in the appendix. We see that the number of K22s per user varies between 0 and \num{7.5e14} for the closed K22s and between 0 and \num{1.1e17} for the open K22s. Note that, due to the algorithm we used in the computations of the global coefficient, we give here the number of K22s and open K22s in which a user is involved as a {\em top vertex} of the K22s. Another distribution would consider users as bottom users of K22s. \giovanni{Too much for a figure in appendix?}

%max K22 7.5 10^14
%max open K22 1.1 10^17

%\begin{conference}
%\ms
%\end{conference}

\nitbf{Clustering coefficient in the mutual graph.} The mutual graph captures the friendship relationships in the social network. The \ccMUT thus is high ($\ccMUTs=10.7 \%$), as cliques of friends are frequent in Twitter. 
\thibvu{A rallonger ?}
%However, bi-directional links are not exclusively friendship relationships. A large part of them is due to the known policy of following back somebody following us (due to politeness or to the hope of being an incentive for other users to follow us back, increasing our number of followers) \giovanni{I understand politeness, the second effect less}. These (somehow) artificial links do not induce a clustering. \giovanni{I took some time to understand. Detail more.} This may explain why the \ccMUT is lower in Twitter than in other undirected social networks like Facebook in which~\cite{NR} reports a clustering coefficient of 5\%. This is also true for the local clustering coefficient, 24\% vs.  40\%~\cite{myers2014information}. 

%\begin{conference}
%\ms
%\end{conference}

\nitbf{Clustering coefficients in the whole graph.}
%For the directed clustering coefficients, we have that \iccs=3.3, \ccTTs=1.9, \ccCTs=1.7. For the undirected metrics, \ccUTs=0.04, and \ccMUTs=3.6. 
We observe that $\iccs=3.3 \% > \ccTTs=1.9 \% > \ccCTs=1.7 \% > \ccUTs=0.11 \%$. Directed metrics better capture the interest relationships in the TS as \ccUTs is very low. The highest parameter is the \iccs. It confirms the hypothesis of this paper that common interests between two users are better captured by the notion of K22 than by a direct link between these users. As expected, the second parameter is the one using transitive triangles. Indeed, they capture a natural way for a user of finding a new interesting user, that is, considering the followings  %\giovanni{better followees?}
of a following, especially after having seen retweets. A bit surprisingly, the \ccCTs is not very low. In fact, a large fraction of the cyclic triangles are explained by corresponding triangles in the mutual graph (triangles of bi-directional links).
\\
A way to artificially take off the social influence in order to focus exclusively on the directed interest part of the graph is to remove the (open and closed) triangles and K22s contained in the mutual graph from the total count. Indeed, each undirected triangle of the mutual graph induces two cyclic triangles and four transitive triangles, and each undirected open triangle induces two open triangles. In the same way, each undirected K22 induces two K22s and each undirected open K22 induces two open K22s. The obtained results are shown in Table~\ref{tab:nomutual}. If we take off those mutual triangles, both the \ccTTs and the \ccCTs values drop to $0.51\%$ and $0.24\%$, respectively, while the \iccs stays about the same at $3.1\%$. This tends to confirm the hypothesis that the directed triangle clusterings somehow measure the friendship part of the \tss more than the interest part. 

\begin{journal}
\noindent
We can even go one step further by computing the number of triangles in the graph in which all bidirectional edges have been removed. In that case, the $\ccCTs$ drastically drops to 0 (we found no cyclic triangles without at least a bidirectional arc in the dataset!) while \ccTTs and \icc stay almost the sames, 3.6 and 4.2 respectively. This confirms that cyclic triangles are artificially created by friendship relations and that the $\ccCTs$ gives no information about the directed part of the graph.
\end{journal}

% In fact, a very large fraction of the cyclic triangles are explained by corresponding triangles in the mutual graph (triangles of bi-directional links): \SI{630}{M} over \SI{725}{M}, that is, 87\%. Indeed, each undirected triangle of the mutual graph induces two cyclic triangles. Because of this, the \ccCTs somehow measures the friendship part of the \tss more than the interest part. 

\begin{table}[t]
\begin{center}
    $
	\begin{array}{|l|cccc|}
	\hline
	& icc & tcc & ccc & ucc \\
	\hline
	Twitter & 3.1\% & 0.51\% & 0.24\% & 0.057\% \\
	\hline
	\end{array}
	$
\caption{Clustering coefficients without the mutual structures. }
\label{tab:nomutual}
\end{center}
%\vspace{-1.cm}
\end{table}

%\rejectforwg{
\begin{figure}
\centering
\includegraphics[width=0.9\linewidth]{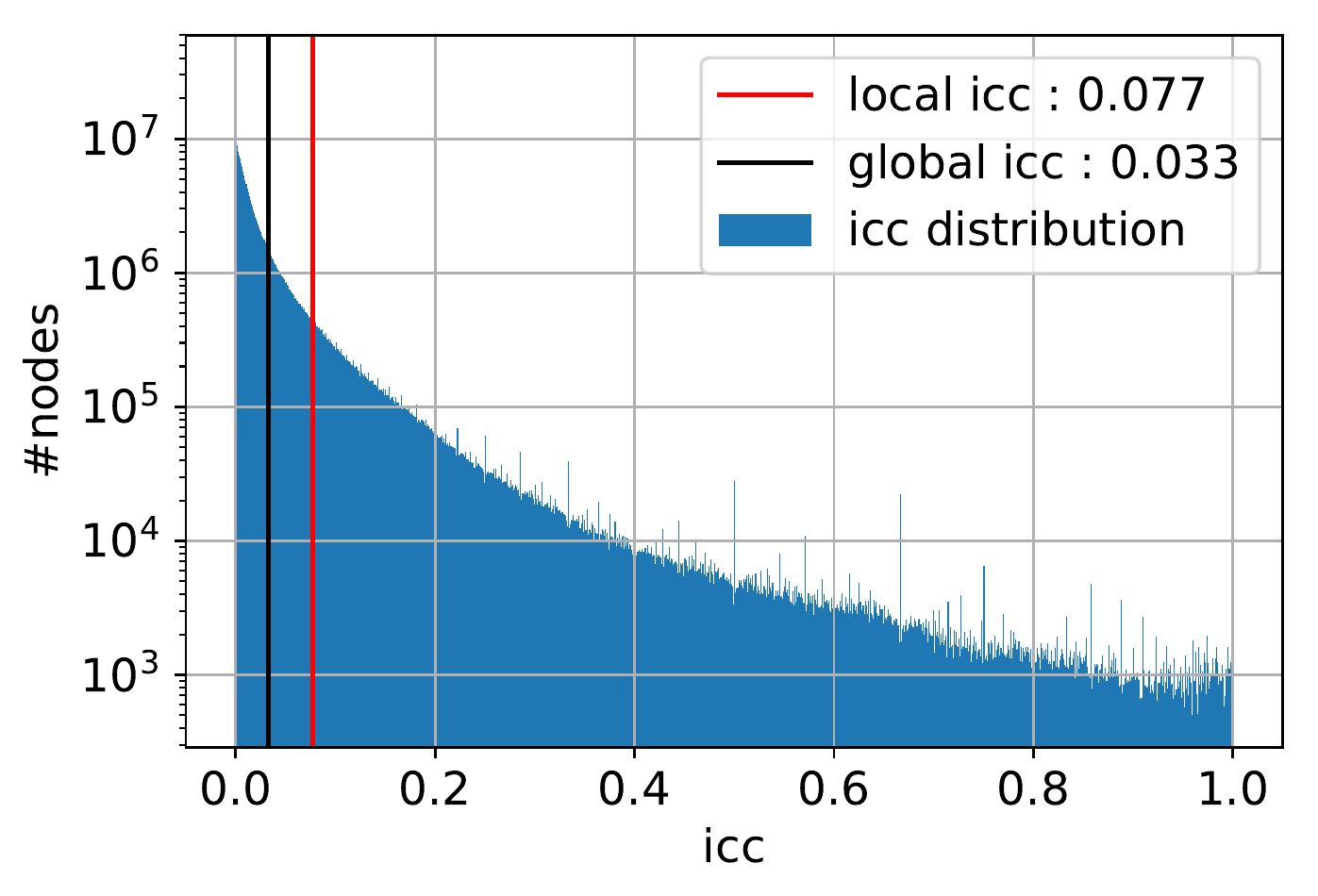}
\caption{Histogram of the distribution of the \icc over all users of the \ts. The vertical bars indicate the value of the glocal \iccs (3.3\%) and the average value (7.7\%) or local \iccs.} %\fred{add the average (0.077) and Twitter global icc ? (=0.033) as a vertical line.}}
\label{fig:Distrib_cc}
%\vspace{-.4cm}
\end{figure}

\ms

\nitbf{Distribution of the \iccs and local clustering.} We also provide the distribution of the values of the \icc over all users (having open K22s) in Figure~\ref{fig:Distrib_cc}. We see that the \iccs greatly varies between 0 and 1. A large number of nodes have a low value of \iccs, e.g., $\num{2.23e7}$ users ($10.2\%$ of the users with open K22s) have a value of 0, meaning they are part of open K22s but not of K22s. At the opposite end, $\num{2.4e4}$ users ($0.011\%$ of the users with open K22s) have a value of 1, meaning that all their open K22s belong to a K22.
%K22s are closed.
The {\em average value} is equal to 7.7\%. This value could be used as a definition of a {\em local $\iccs^{\top}$.} 
%\giovanni{better to introduce them by words first, and then provide the symbols.} \thib{(vrai, mais je n'ai pas vu comment faire autrement... vu qu'on ne l'utilise que là, peut etre pas tres grave.)}
Indeed, as discussed above, the number of K22s and open K22s per user have been computed while considering a user as a top vertex. A second local coefficient, $\iccs_{\bot}$, can be defined for bottom vertices. 
\\
Similarly to what was found in Facebook, the local coefficient has a larger value than the global one. %A reason for this may be that
This may be due to the fact that 
a large number of nodes with few K22s and open K22s (usually nodes with small degrees) only are in a single small strongly connected community, and thus have a higher than average \iccs. On the contrary, a small number of nodes with larger degrees and larger number of K22s and open K22s may be in different communities, leading to smaller than average \iccs. 
\noindent

\section{Results: Other Directed Datasets}
\label{sec:Other-Datasets}

\begin{table*}
  \centering
%  \begin{subtable}[t]{.52\linewidth}
%    \centering
    \begin{tabular}{|l|l|ccc|}
%    \begin{center}
%        $
%    	\begin{array}{|l|l|ccc|}
    	\hline
    	& Is a Social Network & N & $|E|$ & $\frac{|E|_m}{|E|}$ \\
    	\hline
    	Instagram 
    	& Yes
    	& \num{4.5e4} & \num{6.7e5} & 11\% 
    	\\
    	Flickr 
    	& Yes
    	& \num{2.3e6} & \num{3.3e7} & 62\% 
    	\\
    	Web (.edu) 
    	& No
    	& \num{6.9e5} & \num{7.6e6} & 25\% 
    	\\
    	Citations
    	& No
    	& \num{3.8e6} & \num{1.7e7} & 0\% 
    	\\
    	\hline
%    	\end{array}
%    	$
%    \end{center}
    \end{tabular}
    \caption{Datasets information. $N$ is the number of nodes, $|E|$ the number of edges, and $\frac{|E|_m}{|E|}$ the fraction of edges implied in a bidirectional link.}
    \label{tab:other-datasets-informations}
%  \end{subtable}
\end{table*}
\begin{table*}
  \centering
%\hfill
%  \begin{subtable}[t]{.45\linewidth}
%    \centering
    \begin{tabular}{|l|ccccc|}
	\hline
	& icc & tcc & ccc 
%	& \frac{|E|_{m}}{|E|_{TG}} 
	& mcc & ucc\\
	\hline
	Instagram 
	& 12.0\% & 15.4\% & 3.7\% 
%	& 10.9\%
	& 22.6\% & 4.1\% 
	\\
	Flickr
	& 12.4\% & 12.2\% & 9.3\% 
%	& 62.2\%
	& 13.9\% & 10.8\% 
	\\
	Web (.edu)
	& 46.3\% & 59.6\% & 18.8\% 
%	& 25.0\%
	& 78.5\% & 0.69\% 
	\\
	Citations 
	& 22.3\% & 9.1\% & 0\% 
%	& 0\% 
	& \textit{(none)} & 6.7\% 
	\\
%	Web (Google) 
%	& 15.7\% & 47.7\% & 19.7\% 
%	& 30.7\%
%	& 65.8\% & 5.5\% 
%	\\
%	Amazon (co-purchasing) 
%	& 12.8\% & 38.0\% & 25.8\% 
%	& 55.7\%
%	& 50.7\% & 16.6\% 
%	\\
%	LiveJournal 
%	& 12.3\% & 15.3\% & 11.7\% 
%	& 74.8\%
%	& 14.3\% & 11.8\% 
%	\\
	\hline
%	\end{array}
%	$
    \end{tabular}
\caption{Clustering coefficients of the directed datasets.}
\label{tab:other_datasets_total}
%  \end{subtable}
%\caption{Other directed datasets.}
%  \label{mes-beaux-tableaux}
\end{table*}

We computed the different metrics on four other directed networks: two social networks, a web networks and a citation network. The data information are gathered in Table~\ref{tab:other-datasets-informations}, while the clustering coefficients are reported in Table~\ref{tab:other_datasets_total}. We also computed the values of the clustering coefficients without the mutual structures (not provided here); interestingly, those values are close to the ones on the total graphs. 
\\
We observe that the structure of each dataset is revealed by (the mix of) values of the different clustering coefficients, as discussed below.  
\\
\ms
\\
%\begin{itemize}
\nitbf{Instagram}: Instagram is a photo and video-sharing social network. This dataset was collected by Ferrara et al.~\cite{ferrara2014online} in 2014. The network is close to the Twitter one. Nodes corresponds to the accounts, and there is a link $u \rightarrow v$ if the account u follows the account v. The results are quite similar to what we found for Twitter: the $icc$ and $tcc$ are high and of the same order; the $ccc$ is also high because of the bidirectionnal edges (it drastically drops to 0.06\% when removing those links). 
%\thib{Faux : vrai pour l'antimutual, mais pas quand on retire les mutual structures.}). 
The $mcc$ is the highest value, while the $ucc$ is lower than the others. This confirms that social networks exhibit some common characteristics. %/the same general trend. 
% \fred{Note sure of the following. But with some differences. The $tcc$ is higher than the \iccs in Instagram, on the contrary to Twitter. We think it is because Twitter is used more as an information media than Instagram and Flickr which are used to share photos and videos. This is translated into the fact that Twitter as an \iccs higher than the $tcc$ and a more clustered directed part of the social network.} 
\\
\ms
\\
\nitbf{Flickr}: Flickr is an image and video hosting service, which allows you to follow other people on the plateform to see more easily their content. The dataset was collected in 2008 by Mislove et al.~\cite{mislove-2008-flickr}. This is once again a graph of followers of a directed social network. The values are similar to the previous one but for the $ucc$, which is higher. We can notice that Flickr looks more like a social media than Twitter and Instagram, since there is 62\% of links implied in bidirectional. This explains why the undirected clustering coefficient is not so different from the mutual one .
\\
\ms
\\
\nitbf{Berkley-Stanford.edu web pages}: The dataset was collected in 2002 by Leskovec et al.\cite{leskovec2009community}. The nodes represent the pages from berkely.edu and stanford.edu domains and directed edges represent hyperlinks between them.  The $tcc$, $icc$, and $mcc$ are really high. For the $tcc$, this is due to the very hierarchical structure of the institution web pages. As an example, a researcher will be linking towards his group, laboratory, and university in its website, while the group website is linking to its laboratory and university... This strong structure translates into a high value of the $tcc$. As for the $icc$, research and educational domains form naturally strong communities creating large number of common neighbors for individuals of the same domain, and thus a high $icc$. Groups/teams/departments also constitutes strong social communities, leading to a high $mcc$.   
%We believe this is a consequence of the institutions web pages which are links by almost everyone, creating high degree nodes gathering almost all the K22s and transitive triangles.
% - very hierarchic: so very high tcc. A researcher will have the link towards its group, lab, university on its website. Group to lab and university. And lab to university. 
% - very high icc: very strong communities. 
% - very high mcc: very strong social communities (organized group of people/ research team, departments…)
\\
\ms
\\
\nitbf{Citations}: Collected by Leskovec et al.\cite{leskovec2005graphs}, it includes all citations made by patents granted between 1975 and 1999. This is a good example of information network, giving a high value of \iccs of 22.6\%, while the \ccTTs value is 9.1\%. Indeed, research fields and industry domains are strong communities leading to a high \iccs. Moreover, it is also not rare to cite a patent and its citations (the patent acting as a survey), explaining the \ccTTs value.  
%- Very high icc: as there are strong communities in research papers. 
%- Tcc is high but lower: phenomena. Citing the citation of a paper (which acts as survey). 
Note that there are no cyclic triangles nor bidirectional links, because of the temporal structure of citations - a paper will only cite older papers.
%\end{itemize}
\\
\ms
\\
\nitbf{Takeaways}:
The following takeaways summarize the variety of informations given by the different clustering coefficients:
\begin{itemize} 
\item A high value of \iccs indicates the presence of clusters of interests such as research communities or interest fields. 
\item A high value of \ccTTs{} is the sign of an important  {\em local} phenomena of friends' or acquaintances' recommendations and/or of a high hierarchical structure in the dataset.
\item The \ccCTs{} has no real social meaning. If its value can be high in a directed graph, this is only due to the presence of bidirectional arcs and triangles. The closure of a cyclic triangles is very rare in directed networks with no bidirectionnal edges, confirming the general intuition. 
\item Directed networks have a high $mcc$. Indeed, their bidirectional parts (mutual graph) have strong social communities, leading to a high clustering coefficient.  
\item The ucc is usually significantly lower, showing that the directed part of the network is better understood using directed clustering coefficients. 
\item Directed social networks have similar mixes of values of their undirected and directed clustering coefficients, however, with some notable differences, due to their diverse usages and information. 
    % Twitter is used more as an information media than Instagram and Flickr which are used to share photos and videos. This is translated into the fact that Twitter has a \iccs higher than the $tcc$ and a more clustered directed part of the social network. \thib{Comme tu disais, attention car Twitter a dans l'absolu un icc plus petit qu'Instagram...} Flickr is more used between friends than Instagram, as exhibited by its very high fraction of bidirectional links. As such, its $ucc$ is of similar order than its $mcc$. 
\end{itemize}

\section{Model with addition of K22s}
\label{sec:New_Model}

To model complex networks, a model with a high number of triangles was introduced in~\cite{vazquez2003growing}. In this section, we introduce a new random graph model in which the {\em number of K22s is higher} than classical directed random graphs. The model is based on the model from Bollob\'as et al.~\cite{bollobas2003directed} to which we add what we call a K22 event. A K22 event closes an open K22. The principle is that if a user has a common interest with another user, and if this user has another interest, it has an increased chance to be interested and to follow it. We then show that the in-degree and out-degree distributions of the introduced model follow a power law (as many real networks). Lastly, we exhibit the increase of the \icc of the generated graphs with the probability of a K22 event.

\subsection{Presentation of the model}

We recall here the events defining the classic preferential attachment model of~\cite{bollobas2003directed} and define the K22 event. We start with an initial graph $G_0=(V_0,E_0)$. Then, at each time step t:
\begin{itemize}
\item With a probability (1-p) (\textbf{Bollob\'as et al. event}):
\begin{itemize}
\item With a probability $\alpha$, we add a node u and a link leaving this node and reaching an existing node v chosen with a probability proportional to $d_{in}(v)+\delta_{in}$;
\item With a probability $\beta$, we add a node v and a link reaching this node and leaving an existing node u chosen with a probability proportional to $d_{out}(u)+\delta_{out}$;
\item With a probability $1-\alpha-\beta$, we add an edge between two existing nodes, chosen with probability proportional to $d_{out}(u)+\delta_{out}$ for the leaving node u and $d_{in}(v)+\delta_{in}$ for the reached node v.
\end{itemize}
% \end{itemize}
% \begin{itemize}
\item With a probability $p$ (\textbf{K22 event}):
\begin{itemize}
\item[1)] We choose a random node (called $u_1$) with a probability proportional to its out-degree $d_{out}(u_1)$;
\item[2)] We pick uniformly at random an out-neighbor of the node $u_1$ (called $v_1$); 
\item[3)] We pick uniformly at random an in-neighbor of the node $v_1$ (called $u_2$);
\item[4)] We pick uniformly at random pick an out-neighbor of the node $u_2$ (called $v_2$);
\item[5)] We add a link from $u_1$ to $v_2$.
\end{itemize}
\end{itemize}
The idea of the K22 event is to close an open K22 ; since $u_2$ follows $v_1$ and $v_2$ at the same time, $v_1$ and $v_2$ have a higher probability to be similar, and a person $u_1$ following $v_1$ has a higher chance to be interested in $v_2$. 
\\
Note that it is possible to introduce multiedges with the K22 events. Indeed, to make the problem tractable, we allow $u_1=u_2$ in Step 3), or $v_2=v_1$ in Step 4). In the empirical study, we construct the random graphs with the multiedges and we get rid of them at the end of the constructions. We empirically verify that the multiedges do not impact the results in the end of the section. Indeed, most of them appear for low degree nodes and, thus, they do not affect the tail of the degree distributions. 

% Note that it is likely to introduce a multiedge with the K22 event, since we allow $u_1=u_2$ in the step 3), or $v_2=v_1$ in the step 4). In the empirical study, we let the construction of the graph happen with those multiedges, and get rid of them at the end of the construction. We have seen empirically that this number of multiedges is low enough to not impact the results.
\begin{commentaires}
\thib{Courbe du nb de multiedges en annexe en fonction de p et pour différents $\alpha$ ?} 
\end{commentaires}

\subsection{In-degree and out-degree distributions}

We show in what follows that the in- and out-degree distributions of the introduced model follow power-laws, as most real networks. 
More precisely:
\\
\begin{theorem}
The probability $P(i)$ (resp. $P(o)$) for a node to have in-degree $i$ (resp. out-degree $o$) in the new model is:
\[
P(i) \underset{i>>1}{\sim} i^{-(1+\frac{1}{A})}
\quad \mbox{ and } \quad
P(o) \underset{o>>1}{\sim} o^{-(1+\frac{1}{B})}, 
\]
where $A=p+\frac{(1-p)(1-\beta)}{1+(1-p)(\alpha+\beta) \delta_{in}}$ and $B=p+\frac{(1-p)(1-\alpha)}{1+(1-p)(\alpha+\beta)\delta_{out}}$.
\end{theorem}

\noindent
%%%%%%%%%%%%%%%%%%%%%%%%%%%%%%%%%%%%%%%%%%%%%%%%%%%%%%%%%%%%%%% beginning proof DD
\begin{proof}

We first focus on the in-degree distribution. This result is derived from the equation giving the evolution of the number of nodes of in-degree $i$ as a function of time, sometimes called Master Equation. 
\\
Let $G(t)=(V(t),E(t))$ be the graph obtained at time $t$, and $N(t)=|V(t)|$. 
The number of edges at time $t$ is $|E(t)| = t + |E_0| \approx t$, while the number of nodes is $N(t) = (1-p)(\alpha+\beta) (t+|V_0|) \approx (1-p)(\alpha+\beta) t$ when t is high enough. Hence, the mean in-degree (and out-degree) of the network is $m = \frac{1}{(1-p)(\alpha+\beta)}$.

Let us compute the in-degree distribution. Calling $N(i,t)$ the number of nodes of in-degree $i$ at time $t$, we can write the Master Equation: 
\begin{align*}
&N(i,t+1) - N(i,t) = (1-p)\alpha \delta_{0,i} + (1-p)\beta \delta_{1,i}
\\
&\mbox{     } + (1-p)(1-\beta) \frac{i-1+\delta_{in}}{\sum \limits_{i=0}^{+\infty} N(i,t)(i+\delta_{in})} N(i-1,t)\\ 
&\mbox{     } - (1-p)(1-\beta) \frac{i+\delta_{in}}{\sum \limits_{i=0}^{+\infty} N(i,t)(i+\delta_{in})} N(i,t)
\\
%\end{align*}
%\begin{align*}
&\mbox{     } +
p \frac{i-1}{\sum \limits_{i=0}^{+\infty} N(i,t)i} N(i-1,t) - p \frac{i}{\sum \limits_{i=0}^{+\infty} N(i,t)i} N(i,t)
%\label{Master_Equation}
\end{align*}
% \begin{eqnarray*}
% N(i,t+1) - N(i,t) &=& \alpha \delta_{0,i} + \beta \delta_{1,i}
% \\
% &+& (1-p)(1-\beta) \frac{i-1+\delta_{in}}{\sum \limits_{i=0}^{+\infty} N(i,t)(i+\delta_{in})} N(i-1,t) 
% - (1-p)(1-\beta) \frac{i+\delta_{in}}{\sum \limits_{i=0}^{+\infty} N(i,t)(i+\delta_{in})} N(i,t)
% \\
% &+&
% p \frac{i-1}{\sum \limits_{i=0}^{+\infty} N(i,t)i} N(i-1,t) - p \frac{i}{\sum \limits_{i=0}^{+\infty} N(i,t)i} N(i,t), 
% \end{eqnarray*}
where $\delta_{i,j}$ is the Kronecker delta.
\\
The Master Equation formulates the variation of the number of nodes with degree i between time $i$ and time $i+1$. 
The two first terms on the right hand side correspond to the addition of a new node, with degree 0 or 1 (depending on if we are in the first or second case of the Bollob\'as et al. event). The third and fourth terms are the probabilities that, during the Bollob\'as et al. event, an edge is connected to a node of degree $(i-1)$ or $i$. This would lead to the arrival of a new node of degree $i$, or the loss of one of them. Those events occur with probability $(1-p)(\alpha+(1-\alpha-\beta))$. Finally, the last two terms correspond to the probability that an edge connects a node of degree $(i-1)$ or $i$ during the K22 event.
\\
We now show that the probability to connect to a node ($v_2$) of a given degree after following an open K22 is proportional to the degree of this node. Indeed, the probability to connect to a node ($v_2$) of a given degree after following an open K22 is 
\[
P(x=v_2) = \sum \limits_{y \in N^+(v_2)} P(y=u_2) \times \frac{1}{d_{out}(y)},
\]
where $N^+(v_2)$ is the set of in-neighbors of $v_2$, and $u_2$ is defined in the model.
Using the same reasoning, we have 
\[
P(x=u_2) = \sum \limits_{y \in N^-(u_2)} P(y=v_1) \times \frac{1}{d_{in}(y)}
\]
and
\[
P(x=v_1) = \sum \limits_{y \in N^+(v_1)} P(y=u_1) \times \frac{1}{d_{out}(y)}.
\]
Since $P(y=u_1)=\frac{d_{out}(y)}{t}$, we deduce that 
\[
P(x=v_2) = \frac{d_{in}(x)}{t},
\]
which gives us the expected result.

\noindent Using this property and knowing that \[ \sum \limits_{i=0}^{+\infty} i \cdot N(i,t) = |E(t)| = t \] and \[ \sum \limits_{i=0}^{+\infty} N(i,t) \delta_{in} = \delta_{in} N(t) = (1-p)(\alpha+\beta)\delta_{in}, \] we can rewrite the equation as:
% \begin{eqnarray*}
% N(i,t+1) &=& \alpha \delta_{0,i} + \beta \delta_{1,i} 
% \\
% &+& \Big( p \frac{i-1}{1} + (1-p)(1-\beta) \frac{i-1+\delta_{in}}{1 + (1-p)(\alpha+\beta) \delta_{in}} \Big) \frac{N(i-1,t)}{t} \\
% &-& \Big(1 + \big( p \frac{i}{1} + (1-p)(1-\beta) \frac{i+\delta_{in}}{1 + (1-p)(\alpha+\beta) \delta_{in}} \big)  \frac{1}{t} \Big) N(i,t)
% \end{eqnarray*}
\begin{align*}
&N(i,t+1) = \alpha \delta_{0,i} + \beta \delta_{1,i} 
\\
&\mbox{   }+ \Big( p \frac{i-1}{1} + (1-p)(1-\beta) \frac{i-1+\delta_{in}}{1 + (1-p)(\alpha+\beta) \delta_{in}} \Big) \frac{N(i-1,t)}{t} \\
&\mbox{   }- \Big(1 + \big( p \frac{i}{1} + (1-p)(1-\beta) \frac{i+\delta_{in}}{1 + (1-p)(\alpha+\beta) \delta_{in}} \big)  \frac{1}{t} \Big) N(i,t). 
\end{align*}
% \begin{multline*}
% N(i,t+1) = \alpha \delta_{0,i} + \beta \delta_{1,i} 
% \\
% \mbox{   }+ \Big( p \frac{i-1}{1} + (1-p)(1-\beta) \frac{i-1+\delta_{in}}{1 + (1-p)(\alpha+\beta) \delta_{in}} \Big) \frac{N(i-1,t)}{t} \\
% \mbox{   }- \Big(1 + \big( p \frac{i}{1} + (1-p)(1-\beta) \frac{i+\delta_{in}}{1 + (1-p)(\alpha+\beta) \delta_{in}} \big)  \frac{1}{t} \Big) N(i,t)
% \end{multline*}
Let us call 
\[Z \equiv 1+(1-p)(\alpha+\beta) \delta_{in}. \]
% We can use the following lemma from~\cite{chung2006complex}:
% \begin{lemma}
% Let $(a_t)$, $(b_t)$, $(c_t)$ be three sequences such that $a_{t+1}=(1-\frac{b_t}{t})a_t+c_t$, $\underset{t \rightarrow +\infty}{lim} b_t = b>0$, and $\underset{t \rightarrow +\infty}{lim} c_t = c$. Then $\underset{t \rightarrow +\infty}{lim} \frac{a_t}{t}$ exists and equals $\frac{c}{1+b}$.
% \label{lemma:lim_rec}
% \end{lemma}
We need the following lemma from~\cite{durrett2007random}: 
%, which can be found in \cite{durrett2007random}:
\noindent\begin{lemma}[\cite{durrett2007random}]
If we have an equation of the form :
\[ N(i,t+1) = \Big( 1 - \frac{b(t)}{t} \Big) N(i,t) + g(t) \] 
where $b(t) \rightarrow b$ and $g(t) \rightarrow g$ as $t \rightarrow +\infty$, then 
\[ \frac{N(i,t)}{t} \rightarrow \frac{g}{b+1}. \]
\label{Lemma_Durrett}
\end{lemma}
\noindent Using Lemma~\ref{Lemma_Durrett} and calling $P(i)= \lim \limits_{t \rightarrow +\infty} \frac{N(i,t)}{t}$, we have:
\begin{eqnarray*}
P(i) &= \frac{\big( \frac{(1-p)(1-\beta)}{Z} + p \big) (i-1) + \frac{\delta_{in}}{Z}}{1 +\big( \frac{(1-p)(1-\beta)}{Z} + p \big) i + \frac{\delta_{in}}{Z}} P(i-1).
\end{eqnarray*}
Let us call \[ A \equiv \frac{(1-p)(1-\beta)}{Z}+p. \]
We thus have:
\begin{eqnarray*}
P(i) &= \frac{i - 1 + \frac{\delta_{in}}{ZA}}{i + \frac{\delta_{in}}{ZA} + \frac{1}{A}} P(i-1) \\
&= P(1) \prod \limits_{k=2}^{i} \frac{k - 1 + \frac{\delta_{in}}{ZA}}{k + \frac{\delta_{in}}{ZA} + \frac{1}{A}} \\
&= \frac{\Gamma(i+\frac{\delta_{in}}{ZA}) \Gamma(\frac{1}{A}+\frac{\delta_{in}}{ZA}+2) }{\Gamma(i+\frac{\delta_{in}}{ZA}+\frac{1}{A}+1) \Gamma(\frac{\delta_{in}}{ZA}+1)}. 
\end{eqnarray*}
Leading to
\[
P(i) \underset{i>>1}{\sim} i^{-(1+\frac{1}{A})}. 
\]

The {\em out-degree distribution} calculation follows the same method. The master equation is the same, except that  $\delta_{in}$ and $\beta$ are replaced by $\delta_{out}$ and $\alpha$. The slope of the out-degree distribution is thus: \[ P_{out}(o) \underset{o>>1}{\sim} o^{-(1+\frac{1}{B})}, \mbox{with } B= \frac{(1-p)(1-\alpha)}{1+(1-p)(\alpha+\beta)\delta_{out}}+p. \]

\nitbf{Concentration.} We have studied here the mean of the distributions. We now use the Azuma's inequalities to show the concentration around the mean.
We have the following result~\cite{durrett2007random}: Let $X_t$ be a martingale with $|X_s - X_{s-1}| \leq c$ for $1 \leq s \leq t$. Then: 
\[ 
P(|X_t - X_0| > x) \leq \exp(-x^2/2c^2t). 
\]
Let $Z(i,t)$ be the number of vertices of degree $i$ at time $t$ and let $F_s$ denote the $\sigma$-field generated by the choices up to time $s$. We apply the result to $X_s = E(Z(i,t)|F_s)$. We have that $|X_s-X_{s-1}| \leq 2$. %\fred{I guess the following sentence is enough} 
Indeed, when we add an edge in the network, we affect only the degrees of its two end-vertices.
%\fred{following sentence is useless}
%We can understand it by noticing that whether we attach the edge added at time $s$ to two vertices $(u_1,v_1)$ or $(u_2,v_2)$ will not affect the degrees of other vertices of the networks, or the probabilities they will be chosen later; hence it follows $|X_{s}-X_{s-1}| \leq 2$.
Since $Z(i,0)=E(Z(i,t))$, using the result with $x=\sqrt{t \log(t)}$, we have 
\[ 
P(|Z(i,t) - E(Z(i,t))| > \sqrt{t \log(t)}) \leq t^{-\frac{1}{8}}. 
\]
And hence, $\frac{Z(i,t)}{t} \underset{t \rightarrow +\infty}{\rightarrow} P(i)$ in probability. 
\end{proof}
%%%%%%%%%%%%%%%%%%%%%%%%%%%%%%%%%%%%%%%%%%%%%%%%%%%%%%%%%%%%%%% end proof DD

\vspace{.2cm}

\noindent The degree distributions of the model follow power-laws, with exponents between $-2$ and $-\infty$. We notice that, for $p=0$, we recover the exponents of the Bollob\'as et al. model $-(1 + \frac{1 + (\alpha+\beta) \delta_{in}}{1-\beta})$ and $-(1 + \frac{1 + (\alpha+\beta) \delta_{out}}{1-\alpha})$~\cite{bollobas2003directed}, while, when $p$ goes to $1$, the exponent goes to $-2$.

Note that, similarly to the Bollob\'as et al. model, we cannot generate graphs with any wanted mean-degree and fixed slopes of the power-law. Some constraints exist in order to keep $\delta_{in}>0$ and $\delta_{out}>0$. 
%This is why, in the next experiments, fixing $\alpha, \beta$, the slope, and imposing $\delta_{in}>0$ and $\delta_{out}>0$ restrains the possible values of $p$. 
For instance, with $\alpha=\beta=0.4$ and slopes of $-2.5$ (the values of our experiments), $p$ has to stay in the interval $[\frac{1}{6}, \frac{2}{3}]$.

\ms

\begin{figure}[t]
\vspace{-.4cm}
\centering
\includegraphics[scale=0.45]{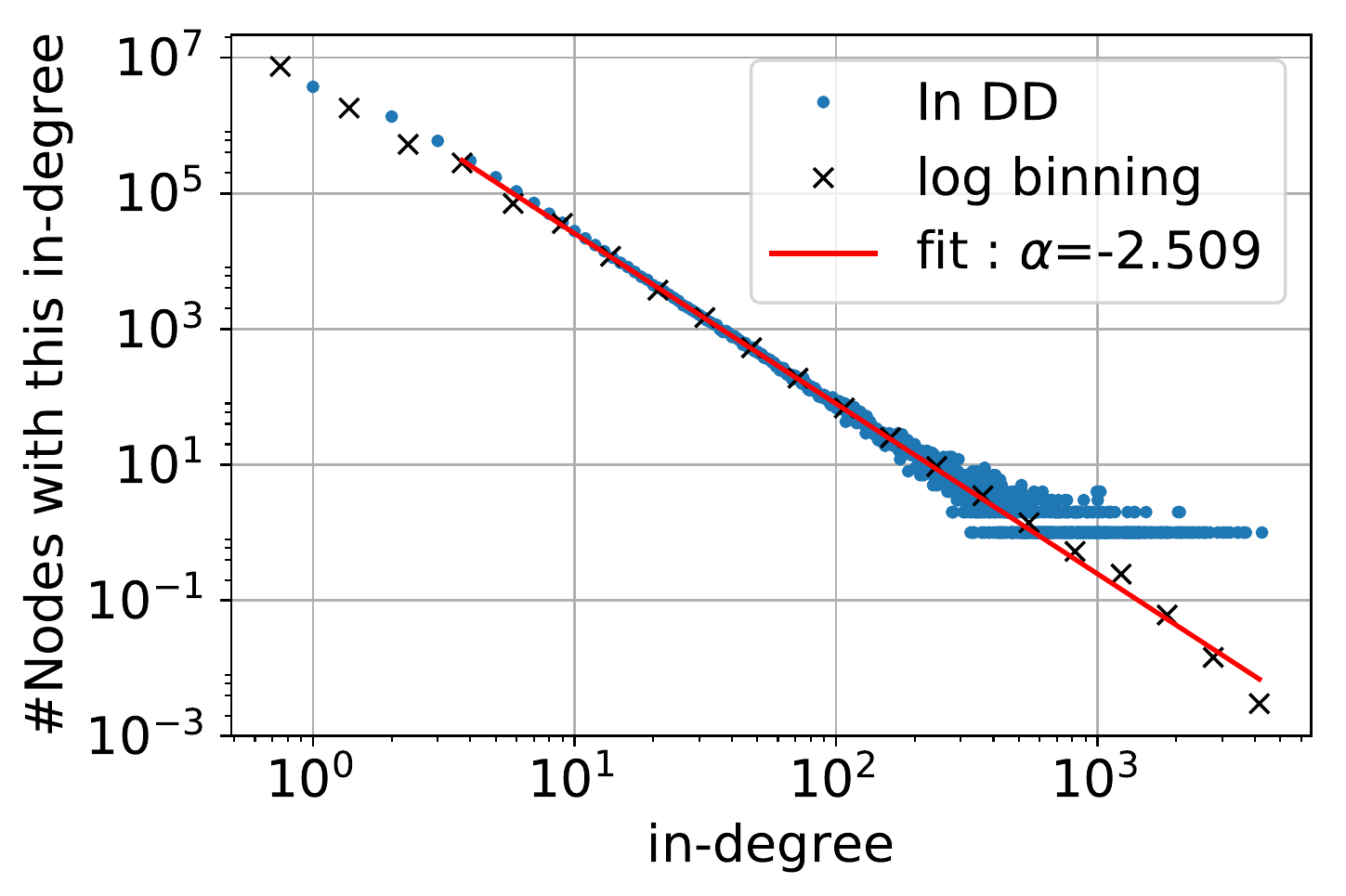}
\includegraphics[scale=0.45]{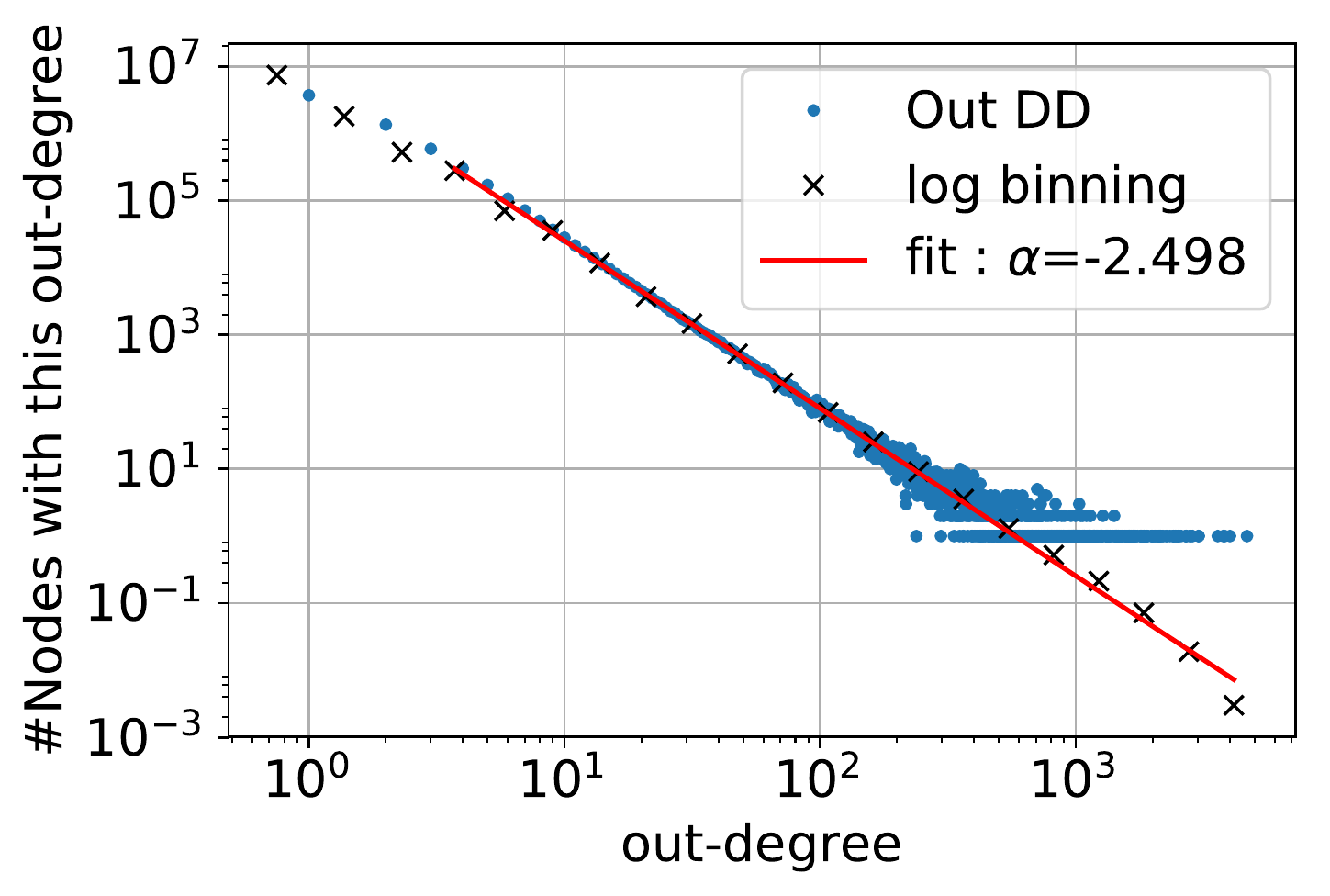}
\caption{In- (Top) and out- (Bottom) degree distributions of a network built with the new model. The obtained distribution is given by the blue points; the black crosses represent the logarithmic binning of the distribution (a mean of a given amount of points on a logarithmic scale). The red straight line is the fit of the logarithmic binning; it has slope of $-2.509$ (resp. $-2.498$) for the in- (resp. out-) degree distribution (expected slopes from analysis are $-2.5$).
\label{fig:new-model-power-laws}}
%\vspace{-.4cm}
\end{figure}

\nitbf{Validation by simulations.} We validate the analysis and the hypothesis by simulation. In Figure~\ref{fig:new-model-power-laws}, we present the in- and out-degree distributions of a network built with our new model as an example. The parameters are fixed to $p=0.5$, $\alpha=\beta=0.4$, and $\delta_{in} = \delta_{out} = 2.0$. In this case, the expected slopes are $-2.5$. The fit is almost perfect: $-2.509$ and $-2.498$ for the in- and out-degree distributions.

\subsection{\Icc of the new model}

\begin{figure}[t]
\vspace{-.4cm}
\centering
\includegraphics[scale=0.5]{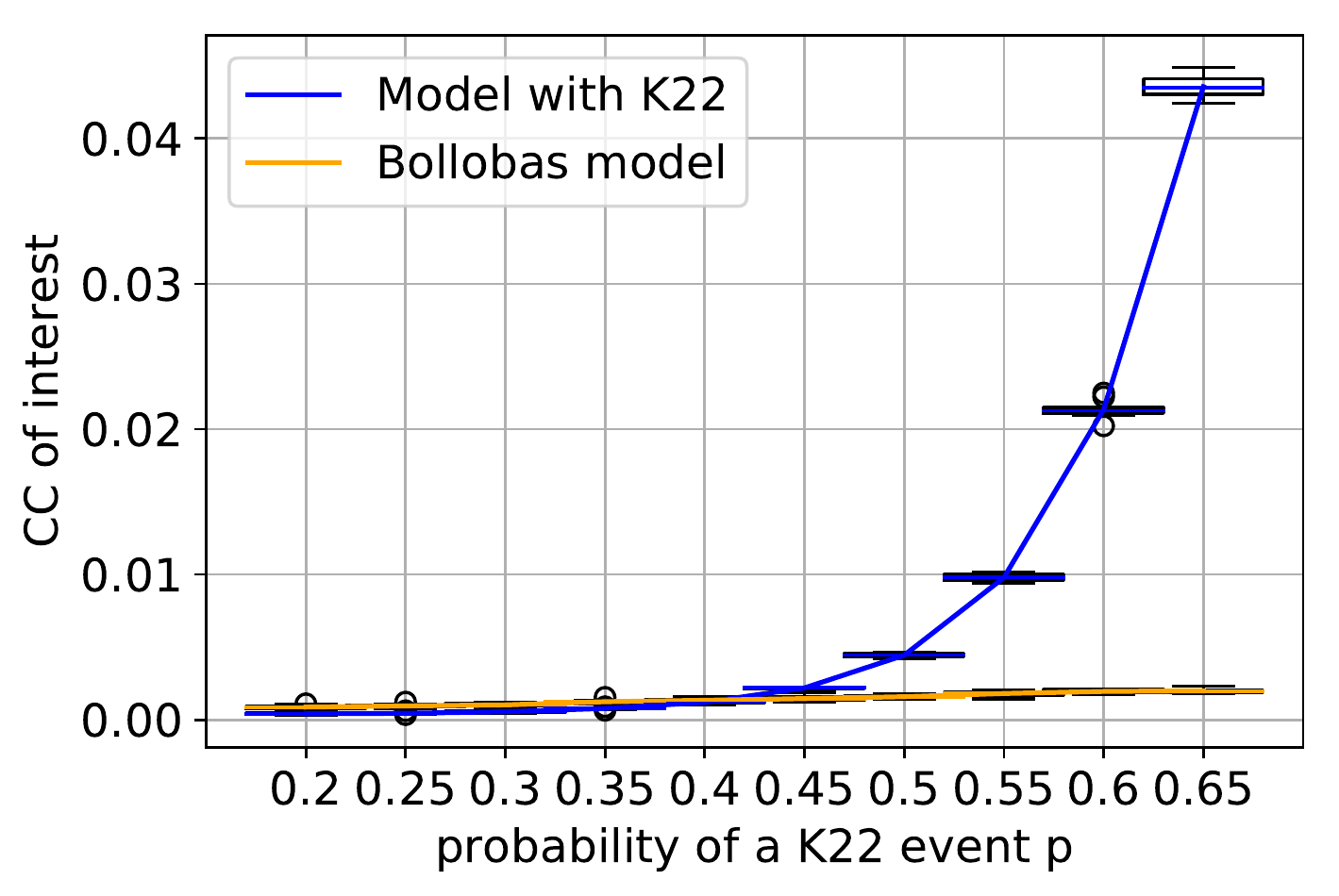}
\caption{\Icc of our new model as a function of $p$, the probability of a K22 event. The value is compared with the one of the Bollob\'as et al. model~\cite{bollobas2003directed}.
%We fixed the slope to $-2.7$, and $\alpha=\beta=0.4$. Each point of the plot is a given value of $p$ in the new model, going from 0 to 0.55 by steps of 0.05 ; at each value of p corresponds a given m, common for the Bollobas and the new model.}
\label{fig:Boxplot_New_Model}
}
%\vspace{-.4cm}
\end{figure}

We show by simulation how the \iccs increases as $p$ increases. We compare it with the one of the Bollob\'as et al. model. Note that, when $p$ increases, the average degree of the model increases. Indeed, the mean degree is $m_{new} = \frac{1}{(1-p)(\alpha+\beta)}$. To compare networks with the same characteristics (mean degrees and exponents of the in-degree distribution), we adapt the parameters of the second model with the value of $p$. 
%To show that the \icc is higher in the introduced model, we compare with the one of the Bollob\'as et al. model using simulations. 
% It looks logical to us to compare the networks of both models with the same characteristics of the final network, i.e. the exponent of the in-degree distribution and the mean degree ; however, the parameters of the model, i.e. $\alpha$, $\beta$, $p$, $\delta_{out}$ and $\delta_{in}$ can be chosen so the two parameters of the network match.
%Knowing that, in the new model, the mean degree is $m_{NM} = \frac{1}{(1-p)(\alpha+\beta)}$, we can see that changing p changes the mean degree. 
\\
Since, in the Bollobas et al. model, the mean degree is $m_{Bol} = \frac{1}{\alpha+\beta}$, we can compare the two models by: choosing the values of $\alpha$, $\beta$, and $p$ for our model. This imposes a value of $m$. We then choose $\alpha$, $\beta$ for the Bollob\'as et al. model, so that the two networks have the same mean degree.
% \begin{itemize}
% \item Choosing a value of $\alpha$, $\beta$ for the new model;
% \item Choosing a value of $p$, which imposes a value of $m$;
% \item Choosing a value of $\alpha$, $\beta$ for the Bollob\'as et al. model, so that the two networks have the same mean degree.
% \end{itemize}
Finally, we choose $\delta_{in}$ so that the exponent of the in-degree distribution stays the same in both networks. In practice, we have fixed the exponent to $-2.5$ and imposed $\alpha_{new} = \beta_{new}=0.4$. 
We compare the \iccs for both models for different values of $p$ and report the results in Figure~\ref{fig:Boxplot_New_Model}. We used graphs of size $N=10^7$ nodes and averaged over 10 networks for each point. We see that the \iccs varies from 0.036\% to 4.4\% when $p$ varies from 0.2 to 0.6.
%\fred{discuss why $p$ no larger than .5 ? }
%\begin{commentaires-imp}
%\thib{To put somewhere else ?}
%\end{commentaires-imp}
%Indeed, the value of the slope is imposing some constraints on the possible values of $p$. As an example, for a slope of $-2.7$, $p$ has to be lower than $0.5$.  However, this is still sufficient to see a substantial increase of the \icc.
% Let us notice that the value of the slope is imposing some constraints : for example, we can't have a slope of $-2.7$, $\delta_{in}>0$ and $p>0.5$ at the same time. Thus for those values, we only did the experiment for $p \in [0,0.5]$.
%(and thus ), or (which is equivalent), for different values of m.

%%%%%%%%%%%%%%%%%%%%%%%%%%%%%%%%%%%%%%%%%%%%%%%%%%%%%%%%

\section{Link Recommendation}
\label{sec:recommendations}

\begin{commentaires-imp}
\fred{could we put the results for 1000 users?}
\end{commentaires-imp}

We propose to use the K22s defined for our metric to carry out link recommendation, as we advocate that the \icc is a good measure of common user interests. For a neighbor, the principle is to recommend links closing open K22s. We define the {\em strength of a link} as the number of open k22s it would close if added to the graph. Links are then recommended by decreasing strengths. Typical recommendation systems propose the strongest link to a user (e.g., Facebook) or a top 10/top 20 list (e.g., Youtube). 

% Links with high strengths are then

We tested our method on the Twitter snapshot. We considered a population of 1000 users selected uniformly at random over the full population of Twitter's users. Note that we excluded users following no one. Indeed, isolated users are not interesting users per se and for this study and they have no TT or K22 recommendations. 

For each node, we computed its open K22s 
(for a node $x$, we follow all its out-neighbors, then for each out-neighbor, we follow its in-neighbors, then for each in-neighbor, we follow its out-neighbors. These last nodes (which were not already followed by $x$) are the recommended nodes. We then count how many times a node is recommended. This gives the link strength.  

We compared the method with classic recommendations using triangles. For example, on Facebook, it is frequent to have a message such as ``8 of your friends know Bob. Do you know Bob?'' Connecting with Bob would close 8 open (undirected) triangles. 
As we are considering a directed graph and are focusing on interest links, we computed recommendations based on transitive triangles, as they have more social sense 
%\fred{pb with the adjective social? (social/friendship vs interest links)} 
than cyclic triangles. For a user $x$, we recommend the out-neighbors of the out-neighbors of $x$. 

the nodes followed by the nodes that $x$ follows are recommended. 
%We computed recommendation based on transitive triangles (nodes followed by nodes that $x$ follows) and undirected triangles. \fred{this one not I guess}
\begin{commentaires}
We could have also studied undirected triangles. Future works?
\end{commentaires}

Note that there are a lot more open K22s than open triangles in the graph, $\num{3.1e18}$ compared to $\num{1.3e14}$. We argue in the following that it allows to make more recommendations and most importantly better recommendations.

% \begin{figure}
% \centering
% \includegraphics[width=.8\columnwidth]{figures/recoms/histogram-max-recom-strength-per-user-bar}
% \caption{Cumulative distribution of the max and 10th TT recommendation strength over 100 random Twitter's users (Left) K22 recommendation (Right) transitive triangle recommendations.}
% \label{fig:top-recom-hists}
% \end{figure}

\begin{figure}
\vspace{-.4cm}
\centering
\includegraphics[width=.48\columnwidth]{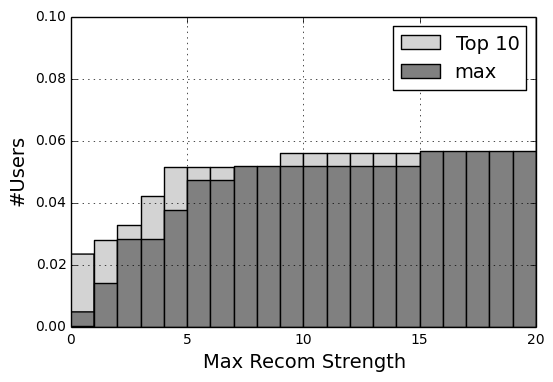}
\includegraphics[width=.48\columnwidth]{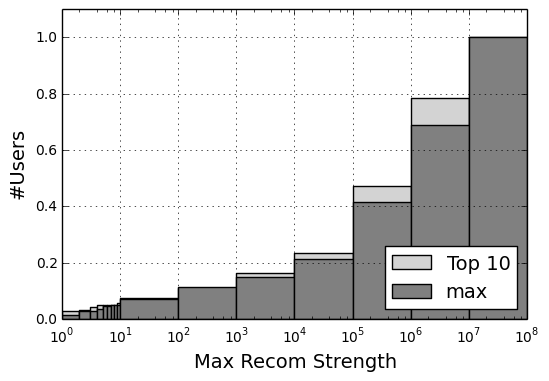}

\includegraphics[width=.48\columnwidth]{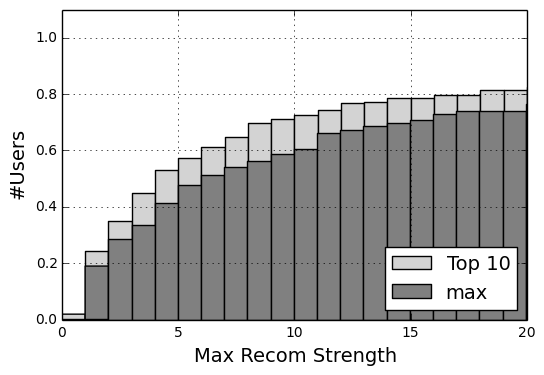}
\includegraphics[width=.48\columnwidth]{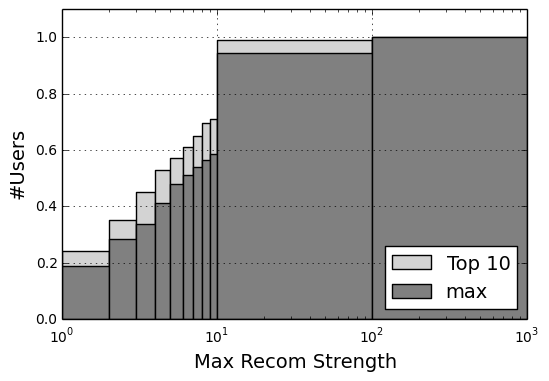}
\vspace{-.2cm}
\caption{Cumulative distribution of the max and 10th recommendation strength over 1000 random Twitter's users for K22 recommendation (Top) and transitive triangle recommendation (Bottom). The left plots are a zoom on recommendations with weak strengths ($\geq 20$). The right plots present the complete cumulative distribution in log scale. 
Beware of the y-scale for the K22 zoom left plot. 
%Beware of the difference in x-scale for the right plots. 
\label{fig:top-recom-hists-random-users}}
\vspace{-.6cm}
\end{figure}

We report in Figure~\ref{fig:top-recom-hists-random-users} histograms of the cumulative distribution over the 1000 random users of the strengths of the recommendation with maximum strength and of the 10th recommendation. The top plots present K22 recommendations while the bottom ones the TT recommendations. The right plots show the complete cumulative distribution in log scale, while the left plots are a zoom on recommendations with weak strengths ($\leq 20$). Beware that the y-scale of the K22 zoom left plot which is between 0 and 0.1. Notice also the difference in x-scale for the right plots. 

\ms

\nitbf{Top/Max recommendation.}
We remark that a small amount of users have TT recommendations and no K22 recommendation. This is due to the fact that for a user with few outgoing links, it is more probable that the followed users are also following at least one other user (providing a TT recommendation) than they are followed by other users (necessary to provide a K22 recommendation). We do not advocate to use only K22 recommendations, but to use it as a complementary tool. In particular, for users with no TT and K22, recommendations would only be made based on global social network statistics (trending topics for example). 

However, when a K22 recommendation exists for a user, it has much more strength than the TT recommendations for her. Indeed, 21\% of users have TT recommendations of strengths 0 or 1. This number is just 1.2\% for K22 recommendations. A recommendation of strength 1 has very good chance to be of no interest, as it is based on the following of a single user over 500 million ones. Similarly, 28\% of users only have TT recommendations of strengths 2 or lower (to be compared with 2.5\% for K22 recommendations). This means that, for a very large portion of users, TT recommendations are based on very few links. On the contrary, more than 94\% of users have a top K22 recommendation with strength more than 10. {\em We are thus able to carry out a meaningful recommendation for the vast majority of users using K22s.} 

\ms

\begin{figure}
\vspace{-.4cm}
\centering
\includegraphics[width=.48\columnwidth]{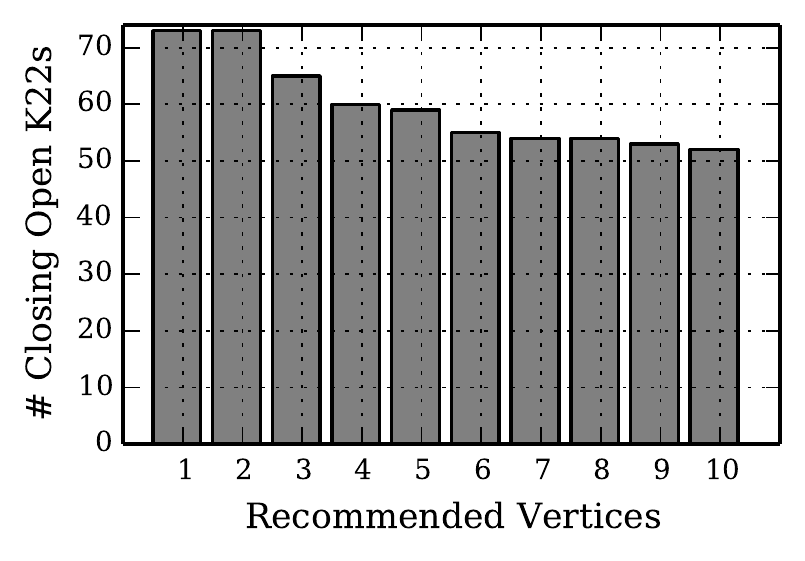}
\includegraphics[width=.48\columnwidth]{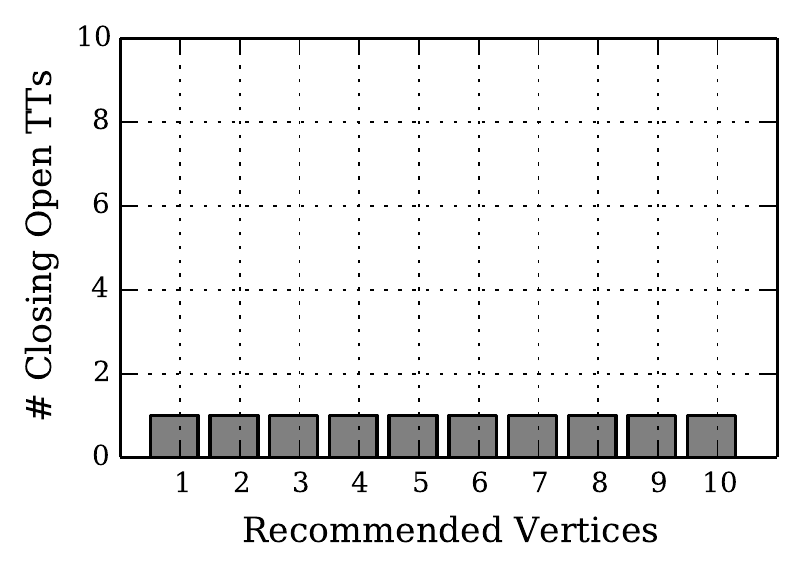}

\includegraphics[width=.48\columnwidth]{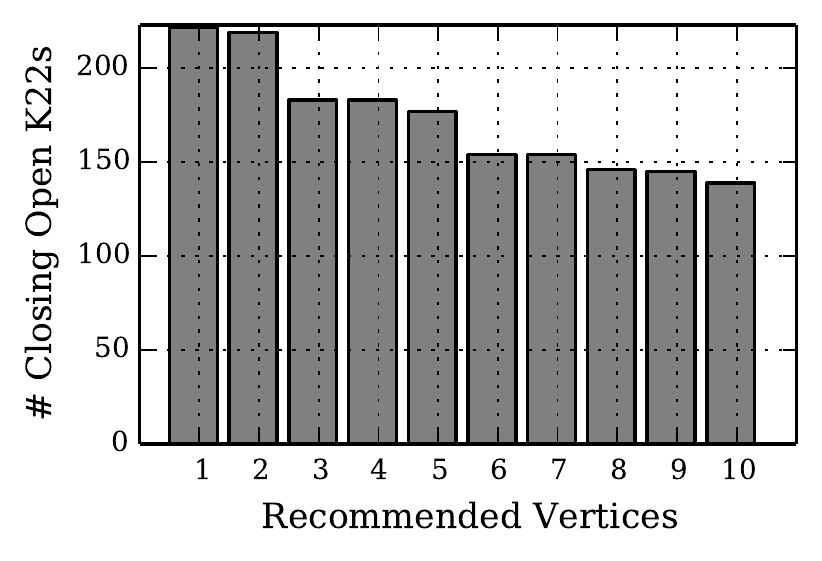}
\includegraphics[width=.48\columnwidth]{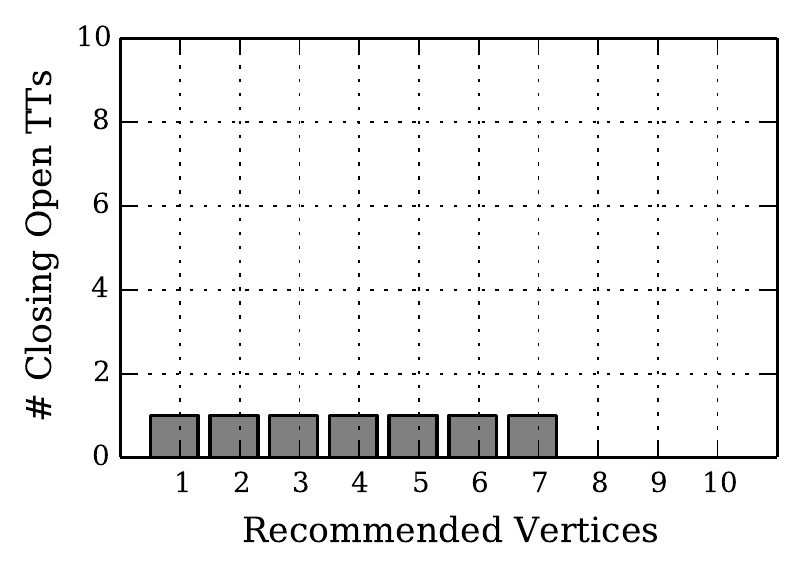}
\vspace{-.4cm}
\caption{Strengths of the top 10 recommendations for 2 typical Twitter users using (Left) K22 recommendations (Right) TT recommendations.}
\label{fig:recom-users}
\vspace{-.5cm}
\end{figure}

\nitbf{Top 10 recommendations.} 
When considering a recommendation system proposing a top 10, we see that 25\% of users have their $10$th TT-recommendation of strength 1 or lower, and 35\% of strength 2 or lower. There does not exist a significant top 10 list for more than one third of users. On the contrary, 94\% of users have their $10$th K22-recommendation with strength higher than 10. Top 10 recommendation systems can thus be implemented for most users using K22s. Moreover, the distribution of recommendation strengths is very flat when using TT (a large number of top recommendations have strength 1), see Figure~\ref{fig:recom-users}. Thus, it is very hard to discriminate between recommended users and to do a meaningful ranking of recommendations. At the opposite end, the distribution usually is steep for K22. It is thus a lot easier to establish a ranking. 

% \begin{figure}
% \centering
% \includegraphics[width=.48\columnwidth]{recom-hist-K22-user24}
% \includegraphics[width=.48\columnwidth]{recom-hist-TT-user24}

% \includegraphics[width=.48\columnwidth]{recom-common-user24}
% \caption{Histograms of the recommendation strengths for a user. (Left) K22 recommendation (Right) transitive triangle recommendations. (Bottom) Common Recommendations.}
% \label{fig:recom-hists}
% \end{figure}

\ms 

\nitbf{Typical users.} We present in Figure~\ref{fig:recom-users} the strengths of the top 10 recommendations using K22 (Left) and TT (Right) for two typical users. For the first one (Top), it is implicated in around 200 triangles, representing each a potential recommendation. However, the strength of the recommendations is very low, just 1 for all of them.  
%the max one is 2 and all the other are one. 
Recommendations for this user would be very bad for two reasons: first, they are based on the choice of only 1  user. Second, if the recommendation system had to propose a top 10, how would it discriminate between the 200 similar potential ones with similar strength. On the contrary, the K22 recommendations have much more strengths: 72 for the 1st and the 2d ones, and 52 for the 10th one. The K22 recommendations are thus much more well-grounded. 
For the second user (Bottom), we observe a similar phenomenon, but with fewer recommendations. It is not even possible to build a top 10 for her using TT as only 8 links can be proposed, and not with a high confidence (strength 1). Conversely, the top 10 K22 recommendations have strengths between 215 and 135.

\begin{commentaires}
The study should be dealt with more depth but we obtain some preliminary results: 
\begin{itemize}
\item Some nodes with no classic triangle recommendation have K22 recommendations. We exhibit an example in Figure~\ref{fig:recommendations}. In this case, recommendations would be only be made based on global social network statistics (trending topics for example). 
Using K22 recommendations thus would allow to make more personal recommendations than a recommend global recommendation. 
\item Note that there are more open K22s than open triangles in the graph $3.1 10^{18}$ compared to $1.3 10^{14}$. This allows to make more recommendations and (better) recommendations. To discuss more...
When no triangle recommendation and more data when there are both (user 3)... A large average number per user, but it is really skewed...
\item The reverse also happens. A node with no K22 recommendation which has triangle recommendations. We do not advocate to use only K22 recommendations, but to use it as a complementary tool. 
\item A large number of nodes do not have undirected mutual recommendations. 
\item Nodes recommended often differs. 
\item 
\end{itemize}
\end{commentaires}

\begin{commentaires}
\begin{figure}
\centering
\includegraphics[width=\columnwidth]{figures/results-by-nodes/nbK22Pot_distribution_by_degree}
\includegraphics[width=\columnwidth]{figures/results-by-nodes/nbK22Pot_distribution_by_degree}
\caption{FUTURE WORK? - Distribution of the number of K22s, open K22s and clustering coefficients per degree of users. Histogram and cdf. \fred{curves do not make sense}}
\label{fig:by-nodes/openK22s-distribution}
\end{figure}
\end{commentaires}

\section{Conclusion}

In this paper, we introduce a new metric, the {\em \icc}, to capture the interest phenomena in a directed graph. Indeed, the classical undirected clustering coefficient apprehends the social phenomena that my friends tend to be connected. However, it is not adequate to take into account directed interest links. 
The \icc is based on the idea that, if two people are following a common neighbor, they have a higher chance to have other common neighbors, since they have at least one interest in common. 
\rejectforwg{We defined the \icc as %$\iccs=4 \cdot \text{number of K22s}/\text{number of open K22s}$. 
\[
\iccs = \frac{4 \cdot \text{number of K22s}}{\text{number of open K22s}}
\]
}
We computed this new metric on a network known to be at the same time a social and information media, a snapshot of Twitter from 2012 with 505 million users and 23 billion links. The computation was made on the total graph, giving the exact value of the \icc, and using sampling methods. The value of the \icc of Twitter is around 3.3\%, higher than (undirected and directed) clustering coefficients introduced in the literature and based on triangles, which we also computed on the snapshot. 
This consolidates the idea that Twitter is indeed used as a social and information media, and that the new metric introduced in this paper captures the interest phenomena. We then proposed a new model, building random directed networks with a high value of K22s, and a new method for link recommendation using K22s.
 As a future work, we would like to investigate further
 link recommendation based on the K22 structure defined for the \icc: in particular, it would be interesting to carry out a real-world user case study to investigate if users are more satisfied by such recommendations. 
 %given by the K22 structure rather than by triangles.

%\input{Biblio_WG.tex}
\bibliographystyle{splncs04}
\bibliography{references}

%\section{Annexes}
%\appendix

%\input{Annex_Twitter.tex}

%\input{Annex_Edge_Sample.tex}

%\input{Annex_MC.tex}

%\input{Annex_Other_Datasets.tex}

%\input{Annex_Proof_DD_New_Model}

\end{document}